\documentclass[a4paper,11pt]{article}
\usepackage{jheppub} 
\usepackage[utf8]{inputenc}
\usepackage{physics}
\usepackage{slashed}
\usepackage{caption}
\usepackage{xcolor}
\usepackage{comment}
\usepackage{multirow}
\usepackage{graphics}
\usepackage{float}
\usepackage{cases}
\usepackage{cancel}
\usepackage{soul}
\usepackage{array}
\usepackage{mathtools}  
\usepackage{amsfonts}
\usepackage{hyperref}
\usepackage{amsmath}
\usepackage{amssymb}
\usepackage{tcolorbox}
\usepackage{tikz}     

\title{\boldmath\boldmath A Supertwistor Formalism for $\mathcal{N}=1,2,3,4\;\text{SCFT}_3$}

\author{Aswini Bala, Sachin Jain, Dhruva K.S., Deep Mazumdar, Vibhor Singh and Brijesh Thakkar}

\affiliation{Indian Institute of Science Education and Research,\\ Dr Homi Bhabha Road, Pashan, Pune, India}

\emailAdd{aswini.bala@students.iiserpune.ac.in}
\emailAdd{sachin.jain@iiserpune.ac.in}
\emailAdd{k.s.dhruva@students.iiserpune.ac.in}
\emailAdd{deepkamal.mazumdar@students.iiserpune.ac.in}
\emailAdd{singh.vibhorrajesh@students.iiserpune.ac.in}
\emailAdd{brijesh.thakkar@students.iiserpune.ac.in}

\abstract{We develop a manifest supertwistor space formalism for three dimensional $\mathcal{N}=1, 2,3,4$ superconformal field theories. This formalism simultaneously makes manifest the supersymmetry, conformal invariance and conservation. We solve two and three point correlators of (half) integer spin conserved supercurrents using the graded supergroup generators. Apart from the superconformal generators, we find that the superhelicity operators are necessary to fix their functional form in this setup. The superhelicity operators can be recast into first order Euler equations which besides the standard polynomial solutions, also admit weak solutions that are distributional in nature. They play an important role in the case of three point functions, where the super-correlator takes the form of a product of delta functions. Interestingly, we find that these super-correlators are extremely simple, elegant and uniform for all spins and for all $\mathcal{N}\leq4$, which resemble the non supersymmetric correlators where the twistors are replaced with appropriately defined supertwistors.}

\begin{document}
\maketitle

\section{Introduction}
Scattering amplitudes serve as the mathematical framework for studying physical observables. Traditionally, they have been studied in momentum space, but the advent of spinor-helicity variables not only has made computations drastically simpler, but also paved the way for remarkable results such as the Parke-Taylor formula \cite{PhysRevLett.56.2459} and the BCFW recursion relation \cite{Britto:2005fq}. Since the inception of twistors in \cite{Penrose:1967wn} and supertwistors in \cite{Ferber:1977qx}, more recently, a great emphasis has been put on understanding scattering amplitudes in terms of twistors \cite{Nair:1988bq,Witten:2003nn,Mason:2009sa,Arkani-Hamed:2009hub} that render a geometric picture to it, for instance \cite{Arkani-Hamed:2012zlh,Arkani-Hamed:2013jha}.
 
Given these developments in scattering amplitudes, one might wonder if such techniques can be used for Conformal Field Theories (CFTs). Traditionally, CFT correlators were computed in the position space \cite{Belavin:1984vu,Rattazzi:2008pe}. However, studying them in momentum space \cite{Maldacena:2011nz,McFadden:2011kk,Ghosh:2014kba,Coriano:2013jba,Bzowski:2013sza,Bzowski:2015pba,Bzowski:2017poo,Farrow:2018yni,Bzowski:2018fql,Bautista:2019qxj,Lipstein:2019mpu,Baumann:2020dch,Jain:2020rmw,Jain:2020puw,Jain:2021wyn,Baumann:2021fxj,Jain:2021qcl,Jain:2021vrv,Jain:2021gwa,Jain:2021whr,Isono:2019ihz,Gillioz:2019lgs,Baumann:2019oyu} has led to interesting results that are not obvious from the perspective of position space, such as the relation between parity even and parity odd correlators \cite{Caron-Huot:2021kjy,Jain:2021wyn,Jain:2021gwa,Jain:2021whr,Skvortsov:2018uru}
and double copy relations \cite{Lipstein:2019mpu,Jain:2021qcl}. Recently, substantial computational progress has been made by studying three-dimensional CFT correlators using spinor-helicity variables \cite{Maldacena:2011nz,Baumann:2020dch,Baumann:2021fxj,Jain:2021vrv,Jain:2024bza}. Given the relationship between spinor helicity and twistors, it is natural to investigate CFT correlators using the latter variables. Although prior work includes the study of three-dimensional scattering amplitudes using twistors \cite{Lipstein:2012kd,Adamo:2017xaf}, an off-shell formalism is required for the study of correlation functions. Recently, \cite{Baumann:2024ttn} developed twistors using the projective rescaling invariance of the Penrose transform and its relation to the embedding space to investigate CFT correlators. A twistor space formalism for conformal Ward identities was developed in \cite{Bala:2025gmz} to solve the CFT correlators, and a Grassmann twistor variable for SCFTs was earlier developed in \cite{Jain:2023idr}.

Supersymmetry plays a crucial role as a theoretical framework. Theories like $\mathcal{N}=4$ super Yang-Mills theory in four dimensions serve as a mathematical laboratory and have paved the way to understand several physical aspects.  Although significant progress has been made for superamplitudes \cite{Elvang:2013cua}, including cases without maximal supersymmetry
\cite{Elvang:2011fx},
It would  be interesting to develop similar techniques for the correlators of operators in superconformal field theories (SCFTs). SCFT correlators are studied primarily in position superspace \cite{Osborn:1998qu,Park:1999pd,Park:1999cw,Nizami:2013tpa,Buchbinder:2021gwu,Buchbinder:2021izb,Buchbinder:2021kjk,Buchbinder:2021qlb,Jain:2022izp,Buchbinder:2023fqv,Buchbinder:2023ndg}, where the super-correlators are constructed using superconformal invariant building blocks. Although this framework has its merits, the analysis becomes technically challenging and is progressively complicated for extended supersymmetries. To alleviate this, there have been some efforts to tackle SCFT correlators using modern amplitude-like techniques such as Grassmann twistors \cite{Jain:2023idr} in recent times.

In this work, we develop a supertwistor space formalism. We setup the supertwistor space for $\mathcal{N}=1,2,3,4$ superconformal field theories in three dimensions and then solve two- and three-point super-correlators of (half) integer spin conserved supercurrents using the graded supergroup generators. These super-correlators  take an extremely simple form, which is reminiscent of twistor space correlators. The outline of this paper is as follows.

\subsection*{Outline}
In section \ref{Result}, we introduce the supertwistor formalism using the supergroup perspective and present the key results of this paper. We then develop this supertwistor space starting from momentum superspace in section \ref{Supertwistor}. We solve the super-correlators of conserved supercurrents for $\mathcal{N}=1$ in section \ref{N1} and then perform the same for extended supersymmetries in section \ref{Extended}. We then solve these super-correlators in manifest supertwistor space in section \ref{sec:ManifestBootstrap} using the graded superconformal generators and helicity operator, which makes the analysis of previous sections much more natural. 
Finally, we summarize and discuss a number of interesting future directions in section \ref{Discussion}.

We supplement this paper with some essential appendices. We lay out the notations and conventions in appendix \ref{app:note}. In appendix \ref{app:sup} we present the superconformal generators in supertwistor variables. Appendix \ref{app:gtv} contains the necessary details on the new Grassmann twistor variables. We then present some manipulations for integral representation of delta-function to accommodate Grassmann variables in appendix \ref{app:delta}. In appendix \ref{app:Fourier}, we present a connection between the supertwistor and dual supertwistor variables by a supersymmetric Fourier transform. We then provide some details on solving superconformal Ward identity and helicity differential equation in manifest supertwistor space in appendix \ref{app:WI} and appendix \ref{app:helicity} respectively. In appendix \ref{app:realitysusy}, we breifly discuss the reality conditions and CPT invariance in supertwistor space. 
We lay out the Euclidean momentum superspace results obtained by a super half-Fourier transform of supertwistor correlators in appendix \ref{app:stsh}. Lastly, we present some essential tensor decompositions of R-symmetry indices for extended supersymmetries in appendix \ref{app:tensor}.

\section{Manifest supertwistor space: supergroup $\And$ results}\label{Result}
In this section, we shall briefly introduce a new off-shell supertwistor space formalism to study $\mathcal{N}=1,2,3,4$ superconformal field theories in three dimensions. We set up the generators, the algebra and determine the invariant dot products of the supergroup. This is followed by a summary of the main results of the paper.

\subsection{The supertwistor space}
We construct the supertwistor space for $\mathcal{N}$-extended supersymmetry using the twistors $(W_A,Z^A)$ \cite{Baumann:2024ttn,Bala:2025gmz} along with a vector of Grassmann twistor variables $(\psi^N,\bar{\psi}^N)$ \cite{Jain:2023idr} as,
\begin{align}\label{supertwistorsN}
\mathcal{Z^A}&=Z^A\oplus\psi^N,\notag\\
\mathcal{W_A}&=W_A\oplus\bar{\psi}^N,
\end{align}
where $\mathcal{A}\in\{A,N\}$, with $A$ being a $Sp(4;\mathbb{R})$ fundamental index and $N$ denotes a $O(\mathcal{N})$ R-symmetry index\footnote{In this paper we are restricting ourselves to $\mathcal{N}\leq 4$ supersymmetry.}. Here $\mathcal{Z}$ and $\mathcal{W}$ are respectively called the supertwistor and the dual supertwistor. These are projective coordinates on the supertwistor space $\mathbb{RP}^{3|\mathcal{N}}$ and its dual space, respectively. We shall see that the action of the superconformal generators becomes extremely simple in this language.

We use supergroup arguments to combine all superconformal generators into a single generator. The $\mathcal{N}$-extended superconformal theory in three dimensions enjoys the symmetry under the $OSp(\mathcal{N}|4;\mathbb{R})$ supergroup\footnote{We refer the reader to the excellent and comprehensive book by Siegel, \href{http://insti.physics.sunysb.edu/~siegel/errata.shtml}{FIELDS} for a discussion of supergroups.}. It has $10+\frac{\mathcal{N}(\mathcal{N}-1)}{2}+4\mathcal{N}$ number of generators with 10 conformal generators, $\frac{\mathcal{N}(\mathcal{N}-1)}{2}$ generators of R-symmetry, and $4 \mathcal{N}$ fermionic generators. The invariant super-tensor, $\Omega_{\mathcal{A}\mathcal{B}}$, associated with this group is defined as,
\begin{align}
    \Omega_{\mathcal{A}\mathcal{B}}=\begin{pmatrix}
        0&\delta_a^{b}&0\\
        -\delta^b_{a} & 0&0\\
        0&0&\mathbb{I}_{\mathcal{N}\cross \mathcal{N}}
    \end{pmatrix}=\Omega^{\mathcal{A}\mathcal{B}},
\end{align}
where $\Omega_{\mathcal{A}\mathcal{B}}\Omega^{\mathcal{AC}}=\delta_\mathcal{B}^\mathcal{C}$.
We exploit the fact that the supergroup naturally comes with a grading. We use this idea to compactly write the functional representation of the infinitesimal generators as follows,
\begin{align}{\label{supertwistorgeneratorZ}}
\mathcal{T}^{\mathcal{AB}}(\mathcal{Z}) \equiv\mathcal{Z}^{\mathcal{(A}} \frac{\partial}{\partial\mathcal{Z}_{\mathcal{\mathcal{B}]}}}= \mathcal{Z}^{\mathcal{A}} \frac{\partial}{\partial\mathcal{Z}_{\mathcal{B}}} + (-1)^{\eta_{\mathcal{A}}\eta_{\mathcal{B}}} \mathcal{Z}^{\mathcal{B}} \frac{\partial}{\partial\mathcal{Z}_{\mathcal{A}}},
\end{align}
where $\eta_{\mathcal{A}}$ is the grading of the index $\mathcal{A}$; with $\eta_\mathcal{A}$ = 0 for the indices of bosonic coordinates of $Sp(4;\mathbb{R})$ and $\eta_\mathcal{A}$ =1 for the indices of fermionic coordinates of $O(\mathcal{N})$ respectively. In the matrix form, the generator \eqref{supertwistorgeneratorZ} assembles all the superconformal generators compactly, 
\begin{align}\label{supertwistormatrix}
 \mathcal{T}^{\mathcal{AB}} & =
    \begin{pmatrix}
    -i P^{ab} &\quad  -iM^a_{b} -2i\delta_{b}^{a} D &  \quad\sqrt{2}e^{\frac{i\pi}{4}}Q^{a\, N}   \\
   -iM^a_{b} -2i\delta_{b}^{a} D & -iK^{ab} &  \frac{e^{\frac{i\pi}{4}}}{\sqrt{2}}S_{a}^{\,N} \\ 
    \sqrt{2}e^{\frac{i\pi}{4}}Q^{a\,M}   & \frac{e^{\frac{i\pi}{4}}}{\sqrt{2}}S_{a}^{M}  &  R^{MN} 
    \end{pmatrix}.
\end{align}
Note the graded nature of the generator $\mathcal{T}^{\mathcal{AB}}$. This feature is absent in four dimensions where the generators come without any symmetrization \cite{Elvang:2013cua}. It is similar to the fact that the four-dimensional momentum $p_{a\dot{a}}$ does not have any symmetrization, while its three-dimensional counterpart $p_{ab}$ is symmetric; see Appendix H of \cite{Bala:2025gmz} for instance.

Although the general literature to the best of our knowledge has the (anti)commutation relations for each superconformal generator treated separately, the super-algebra in these variables \eqref{supertwistorsN} can be compactly written using supertwistor generator \eqref{supertwistorgeneratorZ}.
Using the properties of grading carefully, the associated $\mathfrak{osp}(\mathcal{N}|4;\mathbb{R})$ algebra can be obtained to be,
\begin{align}\label{OSpN4algebra}
[\mathcal{T}^{\mathcal{AB}},\mathcal{T}^{\mathcal{CD}})=\Omega^{\mathcal{CB}}\mathcal{T}^{\mathcal{AD}}+(-1)^{\eta_{\mathcal{C}}\eta_{\mathcal{D}}}\Omega^{\mathcal{DB}}\mathcal{T}^{\mathcal{AC}}+ (-1)^{\eta_\mathcal{A}\eta_\mathcal{B}}\Omega^{\mathcal{CA}} \mathcal{T}^{\mathcal{BD}}+(-1)^{\eta_\mathcal{A}\eta_\mathcal{B}+\eta_\mathcal{C}\eta_\mathcal{D}}\Omega^{\mathcal{DA}}\mathcal{T}^{\mathcal{BC}},
\end{align}
where $[\mathcal{T}^{\mathcal{AB}},\mathcal{T}^{\mathcal{CD}})\equiv \mathcal{T}^{\mathcal{AB}}\mathcal{T}^{\mathcal{CD}}-(-1)^{(\eta_{\mathcal{A}}+\eta_{\mathcal{B}})(\eta_{\mathcal{C}}+\eta_{\mathcal{D}})}\mathcal{T}^{\mathcal{CD}}\mathcal{T}^{\mathcal{AB}}$.
Similarly, we write the generators in dual supertwistor as, 
\begin{align}{\label{supertwistorgeneratorW}}
    \tilde{\mathcal{T}}^{\mathcal{AB}}(\mathcal{W}) =\mathcal{W}^{\mathcal{(A}} \frac{\partial}{\partial\mathcal{W}_{\mathcal{\mathcal{B}]}}} = \mathcal{W}^{\mathcal{A}} \frac{\partial}{\partial\mathcal{W}_{\mathcal{B}}} + (-1)^{\eta_{\mathcal{A}}\eta_{\mathcal{B}}} \mathcal{W}^{\mathcal{B}} \frac{\partial}{\partial\mathcal{W}_{\mathcal{A}}}.
\end{align}
The superconformal Ward identity on super-correlators of conserved supercurrents $\mathbf{J}_s$ then reads off as,
\begin{align}\label{manifesttwistorconformalWard}
\langle 0|\cdots [\mathcal{T}^{\mathcal{AB}},\mathbf{J}_s]\cdots|0\rangle=0.
\end{align}
The solutions to \eqref{manifesttwistorconformalWard} are simply functions of the invariants constructed out of the supertwistors $\mathcal{Z}_i$ and $\mathcal{W}_j$.\footnote{There can be other solutions to superconformal Ward identity which do not depend on these invariant dot products. For instance the delta function $\delta^{(4|\mathcal{N})}(\sum_i c_i\mathcal{Z}_i^\mathcal{A})$ is an $OSp(\mathcal{N}|4;\mathbb{R})$ invariant, but does not figure in the analysis. \label{footnote}} The corresponding dot products invariant under the supergroup $OSp(\mathcal{N}|4;\mathbb{R})$ are as follows,\footnote{We define the dot products as $\mathcal{Z}_i \cdot \mathcal{Z}_j\equiv-\mathcal{Z}_i^\mathcal{A}\Omega_{\mathcal{AB}}\mathcal{Z}_j^\mathcal{B}$ and $\mathcal{W}_i \cdot \mathcal{W}_j\equiv\mathcal{W}_{i\mathcal{A}}\Omega^{\mathcal{AB}}\mathcal{W}_{j\mathcal{B}}$. $\mathcal{Z}_i^\mathcal{A}$ and $\mathcal{W}_{j\mathcal{A}}$ are contracted directly. Refer appendix \ref{app:note} for further details.}
\begin{align}\label{supertwistorsdotExt}
\notag \mathcal{Z}_i \cdot \mathcal{Z}_j&= Z_i \cdot Z_j-\delta_{MN}\psi_i^M\psi_{j}^N,\\ 
\notag\mathcal{W}_i \cdot \mathcal{W}_j&= W_i \cdot W_j+\delta_{MN}\bar{\psi}_i^M\bar{\psi}_{j}^N,\\
\notag\mathcal{Z}_i \cdot \mathcal{W}_j&= Z_i \cdot W_j+\delta_{MN}\psi_i^M\bar{\psi}_{j}^N,\\
\mathcal{W}_i\cdot \mathcal{Z}_j&= W_i \cdot Z_j+\delta_{MN}\bar{\psi}_i^M\psi_{j}^N.
\end{align}
In the context of CFTs \cite{Bala:2025gmz} that while conformal invariance in twistor space fixes the argument of solutions, it was the helicity operator that helped fix the functional form of the two and three point functions of conserved currents. We shall emulate the same analysis for super-correlators in detail in section \ref{sec:ManifestBootstrap}.

For the supersymmetric setting, we present a superhelicity operator ($\mathbf{h}_i$) which also takes into account the helicities of the Grassmann variables. Moreover, following \cite{Arkani-Hamed:2013jha}, we take the \textit{natural} choice of expressing positive helicity supercurrents as a function of $\mathcal{Z}$ and the negative helicity supercurrents as a function of $\mathcal{W}$.\footnote{In \cite{Arkani-Hamed:2013jha}, the authors deal with twistor space scattering amplitudes in the context of Klein space. The on-shell superfields containing massless particles have two independent helicities just like the conserved supercurrents that we study in this paper. Thus, we choose similar conventions regarding their representation in twistor space.}
\begin{align}\label{superhelicity1}
    &\mathbf{h}_j\langle\cdots \hat{\mathbf{J}}_{s_j}^{+}(\mathcal{Z}_j)\cdots\rangle=-\frac{1}{2} \big(\mathcal{Z}_{j\mathcal{A}}\frac{\partial}{\partial \mathcal{Z}_{j\mathcal{A}}}+2\big)\langle\cdots \hat{\mathbf{J}}_{s_j}^{+}(\mathcal{Z}_j)\cdots\rangle = +s_{j} \langle\cdots \hat{\mathbf{J}}_{s_j}^{+}(\mathcal{Z}_j)\cdots\rangle, \notag \\
    & \mathbf{h}_j\langle\cdots\hat{\mathbf{J}}_{s_j}^{-}(\mathcal{W}_j)\cdots\rangle=\frac{1}{2} \big(\mathcal{W}_{j\mathcal{A}}\frac{\partial}{\partial \mathcal{W}_{j\mathcal{A}}}+2 \big)\langle\cdots \hat{\mathbf{J}}_{s_j}^{-}(\mathcal{W}_j)\cdots\rangle = -s_{j} \langle\cdots \hat{\mathbf{J}}_{s_j}^{-}(\mathcal{W}_j)\cdots\rangle.
\end{align}
We shall find that when working with half-integer spins, the alternate conjugate representations for currents are convenient. The action of the superhelicity operator on the conjugate representations can be obtained by a super-Fourier transform between $\mathcal{Z}\leftrightarrow\mathcal{W}$ given by,
\begin{align}\label{superfouriertransform}
    f(\mathcal{W}) = \int d^{4|\mathcal{N}}\mathcal{Z} \, e^{-i \mathcal{Z}\cdot\mathcal{W}} \, \tilde{f}(\mathcal{Z}), \qquad \tilde{f}(\mathcal{Z}) = \int d^{4|\mathcal{N}}\mathcal{W} \, e^{i \mathcal{Z}\cdot\mathcal{W}} \, f(\mathcal{W}).
\end{align}
The details of the same are outlined in the appendix \ref{app:Fourier}. For a detailed explanation regarding the explicit form of the measure, please refer appendix \ref{app:realitysusy}. The results for the helicity operator are respectively given by,
\begin{align}\label{superhelicity2}
&\mathbf{h}_j\langle\cdots\hat{\mathbf{J}}_{s_j}^{+}(\mathcal{W}_j)\cdots\rangle=\frac{1}{2} \big(\mathcal{W}_{j\mathcal{A}}\frac{\partial}{\partial \mathcal{W}_{j\mathcal{A}}}+2 - \mathcal{N} \big)\langle\cdots \hat{\mathbf{J}}_{s_j}^{+}(\mathcal{W}_j)\cdots\rangle = +s_{j} \langle\cdots \hat{\mathbf{J}}_{s_j}^{+}(\mathcal{W}_j)\cdots\rangle \notag, \\
&\mathbf{h}_j\langle\cdots\hat{\mathbf{J}}_{s_j}^{-}(\mathcal{Z}_j)\cdots\rangle=-\frac{1}{2} \big(\mathcal{Z}_{j\mathcal{A}}\frac{\partial}{\partial \mathcal{Z}_{j\mathcal{A}}}+2 - \mathcal{N} \big)\langle\cdots \hat{\mathbf{J}}_{s_j}^{-}(\mathcal{Z}_j)\cdots\rangle = -s_{j} \langle\cdots \hat{\mathbf{J}}_{s_j}^{-}(\mathcal{W}_j)\cdots\rangle. 
\end{align}
Given this setup, let us now present the solutions to the superconformal Ward identities that satisfy the helicity identities. 

\subsection{Results}
We present our main results for the two and three point super-correlators in $\mathcal{N}=1,2,3,4$ for conserved supercurrents $\mathbf{J}_s$. We first discuss the results for integer spin supercurrents and then for half-integer spin supercurrents. While there is no difference at the level of two points, the nature of results is fundamentally different for three-point functions. The net integer super-correlators are homogeneous while the net half integer super-correlator are non-homogeneous.\footnote{Please refer the discussion around equation \eqref{CFT3Euclidthreepoint} for a brief explanation.} Thus, we expect the results to reflect this fact and we cannot obtain a result that is valid simultaneously for both. The results below also include the delta function with negative derivatives, where the representation of the delta function is given to negative derivatives \cite{Baumann:2024ttn,Bala:2025gmz}.
\subsubsection{Integer spin supercurrent}
We begin with the integer spin supercurrent $\mathbf{J}_s$.

\subsubsection*{Two point functions}
The two point function in the two independent helicities respectively are as follows,
\begin{align}\label{TwoPointInteger}
\langle 0|\hat{\mathbf{J}}_s^{+}(\mathcal{Z}_1)\hat{\mathbf{J}}_s^{+}(\mathcal{Z}_2)|0\rangle&\propto\frac{1}{(\mathcal{Z}_1\cdot \mathcal{Z}_2)^{2s+2}},\notag\\
\langle 0|\hat{\mathbf{J}}_s^{-}(\mathcal{W}_1)\hat{\mathbf{J}}_s^{-}(\mathcal{W}_2)|0\rangle&\propto\frac{1}{(\mathcal{W}_1\cdot \mathcal{W}_2)^{2s+2}}.
\end{align}
The above results are actually valid for integer and half-integer spins. However, at the level of three points we find it natural to associate $\mathcal{Z/W}$ to +/- helicity respectively for integer spins but the opposite for half integer spins and thus we choose to represent two point functions in the same variables.
\subsubsection*{Three point functions}
We find the following remarkable results for the three point functions,
\begin{align}\label{ThreePointInteger}
\langle 0| \mathbf{\hat{J}_{s_1}^{h_1}}(\mathcal{T}_1)\mathbf{\hat{J}_{s_2}^{h_2}}(\mathcal{T}_2)\mathbf{\hat{J}_{s_3}^{h_3}}(\mathcal{T}_3)|0\rangle\propto\;\delta^{[s_1+s_2-s_3]}(\mathcal{T}_1\cdot \mathcal{T}_2)\delta^{[s_2+s_3-s_1]}(\mathcal{T}_2\cdot \mathcal{T}_3)\delta^{[s_3+s_1-s_2]}(\mathcal{T}_3\cdot \mathcal{T}_1),
\end{align}
where $\mathcal{T}_i=\mathcal{Z}_i~\text{if}~\mathbf{h}_i=+ ~\text{and}~ \mathcal{T}_i=\mathcal{W}_i ~\text{if}~ \mathbf{h}_i=-$. Note importantly that both two and three point results are natural supersymmetric extensions of non supersymmetric results \cite{Baumann:2024ttn,Bala:2025gmz}. 
The proportionality signs in \eqref{TwoPointInteger} and \eqref{ThreePointInteger} are to accomodate both parity even and parity odd solutions. This will be explicitly discussed in section \ref{sec:ManifestBootstrap}.

\subsubsection{Half integer spin supercurrent}
We now move onto the two and three point functions involving conserved supercurrents of half integer spin supercurrent. 
\subsubsection*{Two point functions}
The two point function in (++) and (- -) helicities respectively are as follows,
\begin{align}\label{TwoPointHalf}
\langle 0|\hat{\mathbf{J}}_s^{+}(\mathcal{W}_1)\hat{\mathbf{J}}_s^{+}(\mathcal{W}_2)|0\rangle&\propto\frac{1}{(\mathcal{W}_1\cdot \mathcal{W}_2)^{-2\big(s+\frac{\mathcal{N}}{2}\big)+2}},\notag\\
\langle 0|\hat{\mathbf{J}}_s^{-}(\mathcal{Z}_1)\hat{\mathbf{J}}_s^{-}(\mathcal{Z}_2)|0\rangle&\propto\frac{1}{(\mathcal{Z}_1\cdot \mathcal{Z}_2)^{-2\big(s+\frac{\mathcal{N}}{2}\big)+2}}.
\end{align}
The reason for $s+\frac{\mathcal{N}}{2}$ rather than $s$ in the exponent is due to the form of the action of the helicity operator \eqref{superhelicity2}.
\subsubsection*{Three point functions}
The three point function in all the helicity configurations can be summarized succinctly as,
\begin{align}\label{ThreePointhalfInteger}
\langle 0| \mathbf{\hat{J}_{s_1}^{h_1}}(\mathcal{I}_1)\mathbf{\hat{J}_{s_2}^{h_2}}(\mathcal{I}_2)\mathbf{\hat{J}_{s_3}^{h_3}}(\mathcal{I}_3)|0\rangle\propto \delta^{[-(s_1+s_2-s_3+\frac{\mathcal{N}}{2})]}(\mathcal{I}_1\cdot \mathcal{I}_2)\delta^{[-(s_2+s_3-s_1+\frac{\mathcal{N}}{2})]}(\mathcal{I}_2\cdot \mathcal{I}_3)\delta^{[-(s_3+s_1-s_2+\frac{\mathcal{N}}{2})]}(\mathcal{I}_3\cdot \mathcal{I}_1),
\end{align}
where $\mathcal{I}_i=\mathcal{W}_i~\text{if}~\mathbf{h}_i=+ $ and $\mathcal{I}_i=\mathcal{Z}_i ~\text{if}~ \mathbf{h}_i=-$. This concludes the summary of our key results. Now we shall construct the supertwistor space from the bottom-up starting from the momentum superspace.

\section{Towards supertwistor variables}\label{Supertwistor}
In this section, we construct the off-shell supertwistor space for 3D SCFTs starting from the off-shell momentum superspace. This amounts to performing a \textit{half-Fourier} transform with respect to the bosonic and fermionic coordinates. Combining ideas from \cite{Baumann:2024ttn,Jain:2023idr} and \cite{Bala:2025gmz}, we now present a supertwistor space description. Let us start with the bosonic coordinates first.

\subsection{The bosonic twistors}
A momentum twistor analysis in Lorentzian signature\footnote{The half-Fourier transform is defined for Lorentzian signature where $\lambda$ and $\bar{\lambda}$ are independent, unlike the Euclidean signature where they are Hermitian conjugates of each other.} was developed in \cite{Baumann:2024ttn,Bala:2025gmz}, where the idea is to take the pair $(\lambda,\Bar{\lambda})$ and perform a \textit{half-Fourier} transform with respect to either $\lambda$ or $\Bar{\lambda}$. 
\begin{align}
\textrm{Twistor}:&\quad Z=(\lambda,\bar{\mu}),\notag\\
\textrm{Dual-Twistor}:&\quad W=(\mu,\bar{\lambda}).
\end{align}

It is important to note that these twistors are real projective coordinates of $\mathbb{RP}^{3}$ and its dual space. 
Let us now set up the stage to go into Grassmann twistor variables, starting from the momentum superspace.

\subsection{The fermionic twistors}
Following \cite{Jain:2023idr}, we begin our construction starting with $\mathcal{N}=1$ case and start with the usual momentum superspace coordinates $(p^\mu,\theta^a)$ where $p^\mu$ is the three momentum and $\theta^a$ is a two-component spinor that incorporates the real Grassmann variables. The supersymmetry generator in these variables are as follows,
\begin{align}\label{QMS}
Q_a&=\frac{\partial}{\partial\theta^a}-\frac{1}{2}\theta_b(\slashed p)_a^b,
\end{align}
where $(\slashed p)_a^b = p_{\mu}(\sigma^{\mu})^b_a$ and $(\sigma_\mu)_a^b$ are the three dimensional sigma matrices. The conserved supercurrents are given by, 
\begin{align}\label{JMS}
\mathbf{J}_{s}^{a_1\cdots a_{2s}}(\mathbf{p},\theta)&=J_{s}^{a_1\cdots a_{2s}}(\mathbf{p})+\theta_{b}J_{s+\frac{1}{2}}^{(a_1\cdots a_{2s}b)}(\mathbf{p})+\frac{\theta^2}{4}(\slashed{p})^{a_1}_{b}J_{s}^{a_2\cdots a_{2s}b}(\mathbf{p}).
\end{align}
The supercurrent \eqref{JMS} satisfies the following conservation equation,
\begin{align}
    D_{a_1} \mathbf{J}_{s}^{a_1\cdots a_{2s}}(\mathbf{p},\theta) = 0,
\end{align}
where $D_{a}$ is the supercovariant derivative. We now employ the super spinor-helicity variables \cite{Jain:2023idr},
\begin{align}\label{SHandTheta}
p^\mu=\frac{1}{2}(\sigma^\mu)_b^a \lambda_a\bar{\lambda}^b,\quad\quad
\theta^a=\frac{\bar{\eta}\lambda^a+\eta \bar{\lambda}^a}{2p},
\end{align}
where $\eta,\bar{\eta}$ are Grassmann variables. Hence, the switch from momentum superspace to its spinor-helicity avatar is purely algebraic,
\begin{align}
(p^\mu,\theta^a)\to(\lambda^a,\bar{\lambda}^a,\eta,\bar{\eta}).
\end{align}
The two independent components of the conserved spin $s$ supercurrents \eqref{JMS} in spinor-helicity variables take the following form,
\begin{align}\label{Jsineta}
\mathbf{J}_s^{-}(\lambda,\bar{\lambda},\eta,\bar{\eta})=e^{-\frac{\eta \bar{\eta}}{4}}J_s^{-}(\lambda,\bar{\lambda})+\frac{\bar{\eta}}{2\sqrt{p}}J_{s+\frac{1}{2}}^{-}(\lambda,\bar{\lambda}),\quad
\mathbf{J}_s^{+}(\lambda,\bar{\lambda},\eta,\bar{\eta})=e^{\frac{\eta \,\bar{\eta}}{4}}J_s^{+}(\lambda,\bar{\lambda})+\frac{\eta}{2\sqrt{p}}J_{s+\frac{1}{2}}^{+}(\lambda,\bar{\lambda}).
\end{align}
where, $\mathbf{J}_s^{\pm}(\eta,\bar{\eta}):=\zeta_{a_1}^{\pm}\cdots \zeta_{a_{2s}}^{\pm}\mathbf{J}_{s}^{a_1\cdots a_{2s}}(\eta,\bar{\eta})$ and $\zeta_{a}^{+}=\frac{\bar{\lambda}_a}{\sqrt{p}},\zeta_{a}^{-}=\frac{\lambda_a}{\sqrt{p}}$ are the positive and negative helicity polarization spinors, respectively.

One can perform a \textit{half-Fourier} transform with respect to either $\eta$ or $\bar{\eta}$ for $\mathbf{J}_s^{\pm}$,
\begin{align}\label{GTT}
\mathbf{\Tilde{J}}_s^{\pm}(\lambda,\bar{\lambda},\bar{\chi},\bar{\eta})=\int d\eta~e^{-\frac{\bar{\chi}\eta}{4}}\mathbf{J}_s^{\pm}(\lambda,\bar{\lambda},\eta,\bar{\eta}),\quad\quad
\mathbf{\Tilde{J}}_s^{\pm}(\lambda,\bar{\lambda},\eta,\chi)=\int d\bar{\eta}~e^{-\frac{\chi\bar{\eta}}{4}}\mathbf{J}_s^{\pm}(\lambda,\bar{\lambda},\eta,\bar{\eta}).
\end{align} 
The Grassmann twistor variables can be obtained after a few variable redefinitions. We relegate the details of this to \ref{app:gtv} which in short leads to following change,
\begin{align}\label{fermionicHFT}
(\lambda^a,\bar{\lambda}^a,\eta,\bar{\eta})\longrightarrow{}(\lambda^a,\bar{\lambda}^a,\psi)\quad\textrm{or}\quad(\lambda^a,\bar{\lambda}^a,\eta,\bar{\eta})\longrightarrow{}(\lambda^a,\bar{\lambda}^a,\bar{\psi}).
\end{align}
Let us note that the two Grassmann variables $\eta\;\textrm{and}\;\bar{\eta}$ decreases to either $\psi$ or $\bar{\psi}$. This yields currents that are similar in form to on-shell supercurrents for $\mathcal{N}\leq4$ in four dimensions, see for instance \cite{Elvang:2011fx}.
The two choices in \eqref{fermionicHFT} imply that the conserved supercurrents in these variables take the following form,
\begin{align}\label{GrassmannTwistorSupercurrents}
\mathbf{\hat{J}}_s^{-}(\lambda,\bar{\lambda},\bar{\psi})=J_s^{-}(\lambda,\bar{\lambda})+\frac{e^{\frac{i\pi}{4}}\bar{\psi}}{\sqrt{2p}}J_{s+\frac{1}{2}}^{-}(\lambda,\bar{\lambda}),\quad&\quad
\mathbf{\hat{J}}_s^{+}(\lambda,\bar{\lambda},\bar{\psi})=e^{\frac{i\pi}{4}}\bar{\psi}J_s^{+}(\lambda,\bar{\lambda})+\frac{1}{\sqrt{2p}}J_{s+\frac{1}{2}}^{+}(\lambda,\bar{\lambda}),\notag\\
\mathbf{\hat{J}}_s^{-}(\lambda,\bar{\lambda},\psi)=e^{\frac{i\pi}{4}}\psi J_s^{-}(\lambda,\bar{\lambda})+\frac{1}{\sqrt{2p}}J_{s+\frac{1}{2}}^{-}(\lambda,\bar{\lambda}),\quad&\quad
\mathbf{\hat{J}}_s^{+}(\lambda,\bar{\lambda},\psi)=J_s^{+}(\lambda,\bar{\lambda})+\frac{e^{\frac{i\pi}{4}}\psi}{\sqrt{2p}}J_{s+\frac{1}{2}}^{+}(\lambda,\bar{\lambda}).
\end{align}
We put these ingredients together to combine the bosonic and fermionic twistors in the following sub-section.
\subsection{The supertwistor variables}
As alluded above, the twistor transforms for the bosonic and fermionic coordinates of momentum superspace can be combined to formulate the supertwistor variables.
\subsubsection*{Supertwistors}
Performing a twistor transform of the supercurrents \eqref{GrassmannTwistorSupercurrents}, we identify $\psi$ as the Grassmann twistor variable\footnote{One may opt for the other choice i.e. $(\lambda,\bar{\mu},\bar{\psi})$, but the supersymmetry generator in these variables take a form of second-order mixed differential equation, which is not in the spirit of twistor space.}$^,$\footnote{The rationale behind rescaling the currents to $\Delta=2$ is that they transform under SCT and translation on the same footing.},
\begin{align}\label{SuperTwistorTrans1}
\hat{\mathbf{J}}_s^{\pm}(\lambda,\bar{\mu},\psi)=\int \frac{d^2\bar{\lambda}}{(2\pi)^2}e^{i\bar{\lambda}\cdot \bar{\mu}}\frac{\hat{\mathbf{J}}_s^{\pm}(\lambda,\bar{\lambda},\psi)}{p^{s-1}},
\end{align}
where the Grassmann twistor transform from $(\eta,\bar{\eta})$ to $\psi$ is already done as mentioned in \eqref{fermionicHFT}.
\subsubsection*{Dual supertwistors}
Performing a dual-twistor transform of the supercurrents \eqref{GrassmannTwistorSupercurrents}, we identify $\bar{\psi}$ as the Grassmann dual-twistor variable,
\begin{align}\label{SuperDualTwistorTrans1}
\tilde{\mathbf{J}}_s^{\pm}(\mu,\bar{\lambda},\bar{\psi})=\int \frac{d^2\lambda}{(2\pi)^2}e^{-i\lambda\cdot \mu}\frac{\hat{\mathbf{J}}_s^{\pm}(\lambda,\bar{\lambda},\bar{\psi})}{p^{s-1}},
\end{align}
where the Grassmann twistor transform from $(\eta,\bar{\eta})$ to $\bar{\psi}$ is already done as mentioned in \eqref{fermionicHFT}. With the (dual-)supertwistor variables at our disposal, let us proceed with solving the super-correlators in these variables.

\section{Solving $\mathcal{N}=1$ super-correlators}\label{N1}
Let us begin with the simplest supersymmetric case i.e. the $\mathcal{N}=1$ scenario. The super-correlator contains several component correlators. It was observed in \cite{Bala:2025gmz} that the conformal generators like $K_\mu$ and $D$, along with the helicity operator fixes the component correlators. The action of supersymmetry generator $Q_a$ on the super-correlator then simply fixes the relative coefficients between the component correlators in the super-correlator. While in principle one needs the action of both supersymmetry generators $Q_a$ and $S_a$ to impose superconformal invariance, the algebra (appendix \ref{app:sup}) implies that just imposing $Q_a$ with SCT ($K_{\mu}$) invariance suffices. Let us setup the stage for this analysis in supertwistor variables.
\subsubsection*{Supertwistors}
The conserved supercurrents in the supertwistor variables take the form following \eqref{GrassmannTwistorSupercurrents} and \eqref{SuperTwistorTrans1},
\begin{align}\label{supertwistorcurrent1}
\notag\mathbf{\hat{J}_s^+}(\lambda, \bar{\mu},\psi)=& \left( \hat{J}_s^+ (\lambda, \bar{\mu}) + e^{\frac{\textbf{i}\pi}{4}}\psi\;\hat{J}_{s+\frac{1}{2}}^+(\lambda, \bar{\mu})\right), \\
\mathbf{\hat{J}_s^-}(\lambda, \bar{\mu},\psi)=& \left(e^{\frac{\textbf{i}\pi}{4}}\psi\; \hat{J}_s^- (\lambda, \bar{\mu}) + \hat{J}_{s+\frac{1}{2}}^-(\lambda, \bar{\mu})\right).
\end{align}
The action of supersymmetry generator in this case is as follows\footnote{The twistor representation of the generators can be obtained by performing twistor transformations as mentioned in \eqref{SuperTwistorTrans1}.},
\begin{align}\label{QTwistor}
[Q_{a},\mathbf{\hat{J}_s^\pm}(\lambda, \bar{\mu},\psi)]   &= \frac{e^{-\frac{i\pi}{4}}}{2\sqrt{2}}\Big( \lambda_a \frac{\partial}{\partial \psi} + \psi \frac{\partial}{\partial \bar{\mu}^a}\Big)\mathbf{\hat{J}_s^\pm}(\lambda, \bar{\mu},\psi).
\end{align}

It was observed in \cite{Elvang:2011fx} that the helicity operator features in the on shell superconformal algebra in context of four dimensional superamplitudes. However, although the helicity operator is not part of our three dimensional super conformal algebra, it plays an essential role in fixing the functional form of the correlators \cite{Bala:2025gmz} and thus it shall play an important role for super-correlators as well. With that in mind, let us define the action of helicity operator on supercurrents \eqref{SuperTwistorTrans1} as,
\small
\begin{align}\label{HTwistor}
&\mathbf{h}\langle \cdots \hat{\mathbf{J}}_s^{+}(\lambda,\bar{\mu},\psi)\cdots\rangle=-\frac{1}{2} \bigg( \lambda_a\frac{\partial}{\partial\lambda_a} + \bar{\mu}_a \frac{\partial}{\partial\bar{\mu}_a} + \psi \frac{\partial}{\partial \psi}+2\bigg)\langle \cdots \hat{\mathbf{J}}_s^{+}(\lambda,\bar{\mu},\psi)\cdots \rangle=+s \langle \cdots \hat{\mathbf{J}}_s^{+}(\lambda,\bar{\mu},\psi)\cdots \rangle,\notag\\&\mathbf{h}\langle \cdots \hat{\mathbf{J}}_s^{-}(\lambda,\bar{\mu},\psi)\cdots \rangle=-\frac{1}{2} \bigg( \lambda_a\frac{\partial}{\partial\lambda_a} + \bar{\mu}_a \frac{\partial}{\partial\bar{\mu}_a} + \psi \frac{\partial}{\partial \psi}+2-1\bigg)\langle \cdots \hat{\mathbf{J}}_s^{-}(\lambda,\bar{\mu},\psi)\cdots \rangle=-s \langle \cdots \hat{\mathbf{J}}_s^{-}(\lambda,\bar{\mu},\psi)\cdots \rangle.
\end{align}
\normalsize
Also, note that unlike in spinor helicity variables where the little group scaling and helicity eigenvalue are one and the same, in super twistor variables we obtain the helicity operator by half Fourier transform of its counterpart which brings up extra terms such as -1 and 2 as can be seen in the above equation. Although the action of the helicity operator is valid only on the f($\lambda,\bar{\mu},\psi$), we can still associate little group scaling for supertwistor variables ${\lambda,\bar{\mu},\psi}$ which is equal to $-\frac{1}{2}$. Moreover, the extra term of $-1$ in the second equation of \eqref{HTwistor} is a consequence of the ``unnatural" choice that we made by expressing the negative helicity currents in the twistor space $(\lambda,\Bar{\mu},\psi)$ rather than in the dual twistor space as in \eqref{HDual}.

\subsubsection*{Dual supertwistors} 
The conserved supercurrents in dual supertwistor variables can be obtained following \eqref{GrassmannTwistorSupercurrents} and \eqref{SuperDualTwistorTrans1},
\begin{align}\label{superdualtwistorcurrent1}
\notag\mathbf{\hat{J}_s^+}(\mu,\bar{\lambda},\bar{\psi})=& \left( e^{\frac{\textbf{i}\pi}{4}}\bar{\psi}\;\hat{J}_s^+ (\mu,\bar{\lambda}) + \hat{J}_{s+\frac{1}{2}}^+(\mu,\bar{\lambda}) \right),\\
\mathbf{\hat{J}_s^-}(\mu,\bar{\lambda},\bar{\psi})=& \left( \hat{J}_s^- (\mu,\bar{\lambda}) + e^{\frac{\textbf{i}\pi}{4}}\bar{\psi}\;\hat{J}_{s+\frac{1}{2}}^-(\mu,\bar{\lambda}) \right).
\end{align}
The action of supersymmetry generator is\footnote{The dual-twistor representation of the generators can be obtained by performing dual-twistor transformations as mentioned in \eqref{SuperDualTwistorTrans1}.},
\begin{align}\label{QDualTwistor}
[Q_{a},\mathbf{\hat{J}_s^\pm}(\mu,\bar{\lambda},\bar{\psi})]   &= \frac{e^{-\frac{i\pi}{4}}}{2\sqrt{2}}\Big(\bar{\lambda}_a \frac{\partial}{\partial \bar{\psi}} - \bar{\psi} \frac{\partial}{\partial \mu^a}\Big) \mathbf{\hat{J}_s^\pm}(\lambda, \bar{\mu},\bar{\psi}).
\end{align}

Let us define the action of helicity operator on supercurrents \eqref{SuperDualTwistorTrans1} as,
\small
\begin{align}\label{HDual}
&\mathbf{h}\langle \cdots \hat{\mathbf{J}}_s^{+}(\mu,\bar{\lambda},\bar{\psi})\cdots\rangle=\frac{1}{2} \bigg( \bar{\lambda}_a\frac{\partial}{\partial\bar{\lambda}_a} + \mu_a \frac{\partial}{\partial\mu_a} + \bar{\psi} \frac{\partial}{\partial\bar{\psi}}+2-1\bigg)\langle \cdots \hat{\mathbf{J}}_s^{+}(\mu,\bar{\lambda},\bar{\psi})\cdots \rangle=+s \langle \cdots \hat{\mathbf{J}}_s^{+}(\mu,\bar{\lambda},\bar{\psi})\cdots \rangle,\notag\\
&\mathbf{h}\langle \cdots \hat{\mathbf{J}}_s^{-}(\mu,\bar{\lambda},\bar{\psi})\cdots \rangle=\frac{1}{2} \bigg( 
\bar{\lambda}_a\frac{\partial}{\partial\bar{\lambda}_a} + \mu_a \frac{\partial}{\partial\mu_a} + \bar{\psi} \frac{\partial}{\partial\bar{ \psi}}+2\bigg)\langle \cdots \hat{\mathbf{J}}_s^{-}(\mu,\bar{\lambda},\bar{\psi})\cdots \rangle=-s \langle \cdots \hat{\mathbf{J}}_s^{-}(\mu,\bar{\lambda},\bar{\psi})\cdots \rangle.
\end{align}
\normalsize 
where the little group scaling of the dual supertwistor variables $(\bar{\lambda},\mu,\bar{\psi})$ is $+\frac{1}{2}$. The extra $-1$ in the first equation of \eqref{HDual} can be obtained by performing a full supertwistor transform from the first line of \eqref{HTwistor}. With the supersymmetry generators \eqref{QTwistor} and \eqref{QDualTwistor}, along with the helicity operators \eqref{HTwistor} and \eqref{HDual} in our arsenal, one can solve the supersymmetric correlators for two and three point functions.

Before we begin, we briefly review the general structure of component currents and Wightman functions. In \cite{Bala:2025gmz}, it was shown that there exist three distinct structures for Wightman functions at three points, namely the homogeneous, non-homogeneous, and odd.
\begin{align}\label{CFT3Euclidthreepoint}
    &\langle 0| J_{s_1}J_{s_2}J_{s_3}|0\rangle=c_{s_1s_2s_3}^{(h)}\langle 0|J_{s_1}J_{s_2}J_{s_3}|0\rangle_{h}+c_{s_1s_2s_3}^{({nh})}\langle 0|J_{s_1}J_{s_2}J_{s_3}|0\rangle_{nh}+c_{s_1s_2s_3}^{(odd)}\langle 0|J_{s_1}J_{s_2}J_{s_3}|0\rangle_{odd}.
\end{align}The nomenclature is due to the analytic continuation that relates them to their Euclidean counterparts. For the Euclidean correlators, (non-)homogeneous refers to the correlators with (non-)trivial Ward identity. The odd correlator refers to the parity odd contribution.

Let us solve the two and three point functions for the $\mathcal{N}=1$ scenario, starting with the integer spin supercurrents.

\subsection{Integer spin supercurrent}
Let us consider the super-correlators for conserved supercurrents with integer spins. In \cite{Jain:2023idr}, it was observed for three point functions of conserved supercurrents with integer spins that only the homogeneous part of the correlator is consistent with supersymmetry, while the non-homogeneous contribution of the correlator vanishes. This analysis becomes much more evident in supertwistor variables.
We start our analysis at the level of two points.
\subsubsection{Two point functions}
Let us begin with two point functions, starting with the plus helicity configuration.
\subsubsection*{(++) helicity}
The two point super-correlator for plus helicity in supertwistor space can be written using \eqref{supertwistorcurrent1}\footnote{$\langle0|\mathbf{\hat{J}}_{s_{1}}(Z_{1}) \mathbf{\hat{J}}_{s_{2}}(Z_{2})|0\rangle|_{s_{1} \neq s_{2}} = 0 $ for conserved currents.},
\begin{align}\label{12+old}
\langle0| \mathbf{\hat{J}_{s}^+}(Z_1,\psi_1) \mathbf{\hat{J}_{s}^+}(Z_2,\psi_2)|0\rangle  &= \langle0|\hat{J}_{s}^+(Z_{1})\hat{J}_{s}^+ (Z_{2}) |0\rangle -i\psi_{1}\psi_{2} \langle 0| \hat{J}_{s+\frac{1}{2}}^+(Z_{1})\hat{J}_{s+\frac{1}{2}}^+(Z_{2}) |0\rangle.
\end{align}
Now, using the fact that component correlators are constrained by solving the component level conformal Ward identities along with helicity identity \cite{Bala:2025gmz,Baumann:2024ttn},  \eqref{12+old} is constrained to take the form (see appendix \ref{app:realitysusy} for more details on the spin dependent factors for component correlator),
\begin{align}\label{12+meta}
\langle 0| \mathbf{\hat{J}_{s}^+}(Z_1,\psi_1) \mathbf{\hat{J}_{s}^+}(Z_2,\psi_2)|0\rangle =  \frac{i^{2s+2}c_s}{(Z_{1}\cdot Z_{2})^{2s+2}}-i~\psi_{1}\psi_{2} \frac{i^{2(s+\frac{1}{2})+2}c_{s+\frac{1}{2}}}{(Z_{1}\cdot Z_{2})^{2(s+\frac{1}{2})+2}},
\end{align}
where $c_s$, $c_{s+\frac{1}{2}}$ are normalization constants for two point functions. Applying the supersymmetry generator \eqref{QTwistor} on \eqref{12+meta} further constraints the relative coefficients to be,
\begin{align}
    c_{s+\frac{1}{2}} = 2c_{s}(s+1).
\end{align} 
Thus, \eqref{12+meta} can be rewritten as,
\begin{align}\label{12+new}
\langle 0| \mathbf{\hat{J}_{s}^+}(Z_1,\psi_1) \mathbf{\hat{J}_{s}^+}(Z_2,\psi_2)|0\rangle =  \frac{c_s\textit{i}^{2 s+2}}{(Z_{1}\cdot Z_{2})^{2s+2}}+\psi_{1}\psi_{2} \frac{2 c_s(s+1) \textit{i}^{2s+2}}{(Z_{1}\cdot Z_{2})^{2(s+\frac{1}{2})+2}}.
\end{align}
One remarkable feature that arises now thanks to the particular fixing of coefficient is that the correlator can now be expressed in a compact manner. The form of \eqref{12+new} is suggestive of a binomial expansion and indeed, one can rewrite it in the following form,
\begin{align}\label{12+}
\langle \mathbf{\hat{J}_{s}^+}\left(Z_1,\psi_1\right) \mathbf{\hat{J}_{s}^+}(Z_2,\psi_2)\rangle = \frac{\textit{i}^{2s+2}(c_s^{even}+ic_s^{odd})}{(Z_{1}\cdot Z_{2} - \psi_{1}\psi_{2})^{2s+2}},
\end{align}
where we have relabeled the constant $c_s=(c_s^{even}+ic_s^{odd})$. This is in accordance with the correct results for momentum-space expressions in spinor helicity for parity even and parity odd parts of the correlators \cite{Bala:2025gmz}. Notice that the denominator in \eqref{12+} is quite reminiscent of the usual twistor dot product $Z_1\cdot Z_2$ \cite{Bala:2025gmz,Baumann:2024ttn}, albeit now replaced with supertwistor variables.
\subsubsection*{(- -) helicity}
A similar analysis can be performed for minus helicity which turns out to give, 
\begin{align}\label{12-}
\langle \mathbf{\tilde{J}_{s}^-}(W_1,\bar{\psi}_1) \mathbf{\tilde{J}_{s}^-}(W_2,\bar{\psi}_2)\rangle =  \frac{(-i)^{2s+2}(c_s^{even}-ic_s^{odd})}{(W_{1}\cdot W_{2} + \bar{\psi}_{1}\bar{\psi}_{2})^{2(s+1)}}.
\end{align}  
Yet again, the denominator in \eqref{12-} is precisely the supersymmetric extension of the usual dual-twistor dot product. Moreover, the factor of $(-i)^{2 s+2}$ correctly incorporates properties of the $CPT$ for $(- -)$ helicity. We defer a detailed analysis of this in Appendix \ref{app:realitysusy}. Let us now move on to solve the three point functions for integer spin supercurrents.

\subsubsection{Three point functions}
Consider the three point correlators for conserved supercurrents of integer spins, starting with the $(+++)$ helicity configuration.

\subsubsection*{+++ helicity}
The three point super-correlator can be expanded into its components using the conserved supercurrents\footnote{Note that the arguments $Z_i$ in the component correlators are suppressed for brevity.} \eqref{supertwistorcurrent1},
\begin{align}\label{13+old}
\langle0| \mathbf{\hat{J}_{s_{1}}^+}(Z_1,\psi_1) \mathbf{\hat{J}_{s_{2}}^+}(Z_2,\psi_2) \mathbf{\hat{J}_{s_{3}}^+}(Z_3,\psi_3)|0\rangle &= \notag
\langle0| \hat{J}_{s_{1}}^+ \hat{J}_{s_{2}}^+ \hat{J}_{s_{3}}^+|0\rangle - i\psi_{1}\psi_{2} \langle0| \hat{J}_{s_{1}+\frac{1}{2}}^+ \hat{J}_{s_{2}+\frac{1}{2}}^+ \hat{J}_{s_{3}}^+|0\rangle \\ &-i\psi_{2}\psi_{3} \langle0| \hat{J}_{s_{1}}^+ \hat{J}_{s_{2}+\frac{1}{2}}^+ \hat{J}_{s_{3}+\frac{1}{2}}^+|0\rangle +i\psi_{3}\psi_{1} \langle0| \hat{J}_{s_{1}+\frac{1}{2}}^+ \hat{J}_{s_{2}}^+ \hat{J}_{s_{3}+\frac{1}{2}}^+|0\rangle.
\end{align}
The conformal invariance along with the helicity identity on component correlators \cite{Bala:2025gmz,Baumann:2024ttn} of \eqref{13+old} imposes the following form,
\begin{align}
\notag \langle0| &\mathbf{\hat{J}_{s_{1}}^+}(Z_1,\psi_1) \mathbf{\hat{J}_{s_{2}}^+}(Z_2,\psi_2) \mathbf{\hat{J}_{s_{3}}^+}(Z_3,\psi_3)|0\rangle  \\ \notag  &= c_{s_1,s_2,s_3}~(-\textit{i})^{s_1+s_2+s_3}~\delta^{[s_1+s_2-s_3]}(Z_{1}\cdot Z_{2})\delta^{[s_2+s_3-s_1]}(Z_{2}\cdot Z_{3})\delta^{[s_3+s_1-s_2]}(Z_{3}\cdot Z_{1})  \\ \notag
&-i~(-\textit{i})^{s_1+s_2+s_3+1}c_{s_1+\frac{1}{2},s_2+\frac{1}{2},s_3}\;\psi_{1}\psi_{2}\; \delta^{[s_1+s_2+1-s_3]}(Z_{1}\cdot Z_{2})\delta^{[s_2+s_3-s_1]}(Z_{2}\cdot Z_{3})\delta^{[s_3+s_1-s_2]}(Z_{3}\cdot Z_{1}) \\ \notag 
&-i~(-\textit{i})^{s_1+s_2+s_3+1}c_{s_1,s_2+\frac{1}{2},s_3+\frac{1}{2}}\;\psi_{2}\psi_{3}\; \delta^{[s_1+s_2-s_3]}(Z_{1}\cdot Z_{2})\delta^{[s_2+s_3+1-s_1]}(Z_{2}\cdot Z_{3})\delta^{[s_3+s_1-s_2]}(Z_{3}\cdot Z_{1}) \\
&+i~(-\textit{i})^{s_1+s_2+s_3+1}c_{s_1+\frac{1}{2},s_2,s_3+\frac{1}{2}}\;\psi_{3}\psi_{1}\; \delta^{[s_1+s_2-s_3]}(Z_{1}\cdot Z_{2})\delta^{[s_2+s_3-s_1]}(Z_{2}\cdot Z_{3})\delta^{[s_3+s_1+1-s_2]}(Z_{3}\cdot Z_{1}).  
\end{align}
The action of supersymmetry generator \eqref{QTwistor} on above equation further constraints the relation between the OPE coefficients, thus resulting in,
\begin{align}\label{eqn4.15}
&\notag\langle0| \mathbf{\hat{J}_{s_{1}}^+}(Z_1,\psi_1) \mathbf{\hat{J}_{s_{2}}^+}(Z_2,\psi_2) \mathbf{\hat{J}_{s_{3}}^+}(Z_3,\psi_3)|0\rangle \\ \notag &= c_{s_1,s_2,s_3}(-\textit{i})^{s_1+s_2+s_3}\Big(\delta^{[s_1+s_2-s_3]}(Z_{1}\cdot Z_{2})\delta^{[s_2+s_3-s_1]}(Z_{2}\cdot Z_{3})\delta^{[s_3+s_1-s_2]}(Z_{3}\cdot Z_{1})  \\ \notag
&+\;\psi_{1}\psi_{2}\; \delta^{[s_1+s_2+1-s_3]}(Z_{1}\cdot Z_{2})\delta^{[s_2+s_3-s_1]}(Z_{2}\cdot Z_{3})\delta^{[s_3+s_1-s_2]}(Z_{3}\cdot Z_{1}) \\ \notag 
&+\;\psi_{2}\psi_{3}\; \delta^{[s_1+s_2-s_3]}(Z_{1}\cdot Z_{2})\delta^{[s_2+s_3+1-s_1]}(Z_{2}\cdot Z_{3})\delta^{[s_3+s_1-s_2]}(Z_{3}\cdot Z_{1}) \\
&+\;\psi_{3}\psi_{1}\; \delta^{[s_1+s_2-s_3]}(Z_{1}\cdot Z_{2})\delta^{[s_2+s_3-s_1]}(Z_{2}\cdot Z_{3})\delta^{[s_3+s_1+1-s_2]}(Z_{3}\cdot Z_{1})\Big).  
\end{align}
Using the integral representation of the delta function, the three point super-correlator can be expressed in a compact form, which is reminiscent of the non-supersymmetric case \cite{Baumann:2024ttn,Bala:2025gmz}. We defer the details of this manipulation in \ref{app:delta}.
\begin{align}\label{13+}
\notag\langle0| \mathbf{\hat{J}_{s_{1}}^+}(Z_1,\psi_1) \mathbf{\hat{J}_{s_{2}}^+}(Z_2,\psi_2) \mathbf{\hat{J}_{s_{3}}^+}(Z_3,\psi_3)|0\rangle=& (-\textit{i})^{s_1+s_2+s_3}~(c_{s_1s_2s_3}^{(even)}+ic_{s_1s_2s_3}^{(odd)})~\delta^{[s_1+s_2-s_3]}(Z_{1}\cdot Z_{2} - \psi_{1}\psi_{2})\\
&\delta^{[s_2+s_3-s_1]}(Z_{2}\cdot Z_{3} - \psi_{2}\psi_{3})\delta^{[s_3+s_1-s_2]}(Z_{3}\cdot Z_{1} - \psi_{3}\psi_{1}),
\end{align}
where $c_{s_1,s_2,s_3}=(c_{s_1s_2s_3}^{(even)}+ic_{s_1s_2s_3}^{(odd)})$. One thus obtains the three point super-correlators in terms of the supersymmetric extension of the twistor dot product. Let us now repeat the analysis for a different helicity configuration.

\subsubsection*{++- helicity}
Consider the three point super-correlator of conserved supercurrents with two plus and one minus helicity. Repeating the same analysis as above yields the following result

\footnote{These are the super-correlators that lead to the correct Wightman functions in spinor-helicity variables. It has been shown in \cite{Bala:2025gmz} that performing an inverse half-Fourier transform to the twistor space correlators yields spinor-helicity Wightman functions. One can do the same for the supersymmetric correlators. These Wightman functions are related to their Euclidean correlators in a very subtle way, which is discussed in detail in \cite{Bala:2025gmz}. We enumerate their spinor-helicity counterparts in appendix \ref{app:stsh}.},
\begin{align}\label{13mix}
\langle0| \mathbf{\hat{J}_{s_{1}}^+}(Z_1,\psi_1) \mathbf{\hat{J}_{s_{2}}^+}(Z_2,\psi_2) \mathbf{\tilde{J}_{s_{3}}^-}(W_3,\bar{\psi}_3)|0\rangle=& (-\textit{i})^{s_1+s_2+s_3}~(c_{s_1s_2s_3}^{(even)}+ic_{s_1s_2s_3}^{(odd)})~\delta^{[s_1+s_2-s_3]}(Z_{1}\cdot Z_{2} - \psi_{1}\psi_{2})\notag\\
&\delta^{[s_2+s_3-s_1]}(Z_2 \cdot W_3+\psi_2\bar{\psi}_3)\delta^{[s_3+s_1-s_2]}(W_3\cdot Z_1+\bar{\psi}_3\psi_1).
\end{align}
A similar analysis can be performed for the other helicity configurations as well. One observes that the arguments always depend on the following dot products,
\begin{align}\label{supertwistorsdot1}
\notag \mathcal{Z}_i \cdot \mathcal{Z}_j&= Z_i\cdot Z_j-\psi_i\psi_{j},\\ 
\notag\mathcal{W}_i \cdot \mathcal{W}_j&=W_i\cdot W_j+\bar{\psi}_i\bar{\psi}_{j},\\
\notag\mathcal{Z}_i \cdot \mathcal{W}_j&=Z_i\cdot W_j+\psi_i\bar{\psi}_{j},\\
\mathcal{W}_i\cdot \mathcal{Z}_j&=W_i\cdot Z_j+\bar{\psi}_i\psi_{j}.
\end{align}
Let us now systematically organize the three point Wightman super-correlators, starting with the homogeneous part.
\subsubsection*{Homogeneous correlators}
A key point to note here is the choice of $Z$ for plus helicities and $W$ for minus helicities in the above examples. It has been studied extensively in \cite{Bala:2025gmz} for non supersymmetric case that the Z-twistor gives rise to homogeneous part of the correlators and W-twistor gives rise to non-homogeneous part of the correlators in plus helicity and vice-versa for minus helicity. Thus in all the above cases for the three point-functions, we precisely obtain the homogeneous part of the super-correlators which is in agreement with \cite{Jain:2023idr}. The $(+++)$ helicity answer in \eqref{13+} then reads off as,
\begin{align}\label{inthomo}
\langle0| \mathbf{\hat{J}_{s_{1}}^+}(Z_1,\psi_1) \mathbf{\hat{J}_{s_{2}}^+}(Z_2,\psi_2) \mathbf{\hat{J}_{s_{3}}^+}(Z_3,\psi_3)|0\rangle_h=&(-\textit{i})^{s_1+s_2+s_3}~(c_{s_1s_2s_3}^{(h)}+ic_{s_1s_2s_3}^{odd})~ \notag \\&\delta^{[s_1+s_2-s_3]}(\mathcal{Z}_1 \cdot \mathcal{Z}_2)\delta^{[s_2+s_3-s_1]}(\mathcal{Z}_2 \cdot \mathcal{Z}_3)  \delta^{[s_3+s_1-s_2]}(\mathcal{Z}_3 \cdot \mathcal{Z}_1).
\end{align}
Let us now consider the cases involving the non-homogeneous part of the super-correlators.
\subsubsection*{Non-homogeneous correlators}
Consider a potential non-homogeneous contribution to (+++) helicity. The apt choice of supertwistor variable for this case will be $(W,\bar{\psi})$ based on the discussion in \cite{Bala:2025gmz}. The super-correlator can then be expanded in component correlators using the supercurrent \eqref{superdualtwistorcurrent1} as follows,
\begin{align}\label{13+nhold}
\langle0| \mathbf{\hat{J}_{s_{1}}^+}(W_1,\bar{\psi}_1) \mathbf{\hat{J}_{s_{2}}^+}(W_2,\bar{\psi}_2) \mathbf{\hat{J}_{s_{3}}^+}(W_3,\bar{\psi}_3)|0\rangle_{nh} &= e^{\frac{3i\pi}{4}}\bar{\psi}_1\bar{\psi}_2\bar{\psi}_3\notag
\langle0| \hat{J}_{s_{1}}^+(W_1) \hat{J}_{s_{2}}^+(W_2) \hat{J}_{s_{3}}^+(W_3)|0\rangle_{nh} \\\notag&+ e^{\frac{i\pi}{4}}\bar{\psi}_{3} \langle0| \hat{J}_{s_{1}+\frac{1}{2}}^+(W_1) \hat{J}_{s_{2}+\frac{1}{2}}^+(W_2) \hat{J}_{s_{3}}^+(W_3)|0\rangle_{nh} \\\notag &+e^{\frac{i\pi}{4}}\bar{\psi}_{1} \langle0| \hat{J}_{s_{1}}^+(W_1) \hat{J}_{s_{2}+\frac{1}{2}}^+(W_2) \hat{J}_{s_{3}+\frac{1}{2}}^+(W_3)|0\rangle_{nh} \\
&+e^{\frac{i\pi}{4}}\bar{\psi}_{2} \langle0| \hat{J}_{s_{1}+\frac{1}{2}}^+(W_1) \hat{J}_{s_{2}}^+(W_2) \hat{J}_{s_{3}+\frac{1}{2}}^+(W_3)|0\rangle_{nh}.
\end{align}
Note in particular the above equation is Grassmann odd. Conformal invariance along with helicity identity then constraints the component correlators in \eqref{13+nhold}, resulting in,
\begin{align}\label{13+nh}
\notag&\langle0| \mathbf{\hat{J}_{s_{1}}^+}(W_1,\bar{\psi}_1) \mathbf{\hat{J}_{s_{2}}^+}(W_2,\bar{\psi}_2) \mathbf{\hat{J}_{s_{3}}^+}(W_3,\bar{\psi}_3)|0\rangle_{nh}= e^{\frac{i\pi} {4}} \\\notag
&\Big( i (-\textit{i})^{-s_1-s_2-s_3}c^{(nh)}_{s_1,s_2,s_3}\bar{\psi}_1\bar{\psi}_2\bar{\psi}_3\delta^{[-(s_1+s_2-s_3)]}(W_{1}\cdot W_{2})\delta^{[-(s_2+s_3-s_1)]}(W_{2}\cdot W_{3})\delta^{[-(s_3+s_1-s_2)]}(W_{3}\cdot W_{1})  \\ \notag
&+(-\textit{i})^{-s_1-s_2-s_3-1}c^{(nh)}_{s_1+\frac{1}{2},s_2+\frac{1}{2},s_3}\;\bar{\psi}_{3}\; \delta^{[-(s_1+s_2+1-s_3)]}(W_{1}\cdot W_{2})\delta^{[-(s_2+s_3-s_1)]}(W_{2}\cdot W_{3})\delta^{[-(s_3+s_1-s_2)]}(W_{3}\cdot W_{1}) \\ \notag 
&+(-\textit{i})^{-s_1-s_2-s_3-1}c^{(nh)}_{s_1,s_2+\frac{1}{2},s_3+\frac{1}{2}}\;\bar{\psi}_{1}\; \delta^{[-(s_1+s_2-s_3)]}(W_{1}\cdot W_{2})\delta^{[-(s_2+s_3+1-s_1)]}(W_{2}\cdot W_{3})\delta^{[-(s_3+s_1-s_2)]}(W_{3}\cdot W_{1}) \\
&+(-\textit{i})^{-s_1-s_2-s_3-1}c^{(nh)}_{s_1+\frac{1}{2},s_2,s_3+\frac{1}{2}}\;\bar{\psi}_{2}\; \delta^{[-(s_1+s_2-s_3)]}(W_{1}\cdot W_{2})\delta^{[-(s_2+s_3-s_1)]}(W_{2}\cdot W_{3})\delta^{[-(s_3+s_1+1-s_2)]}(W_{3}\cdot W_{1})\Big). 
\end{align}
Interestingly, the action of supersymmetric generator \eqref{QDualTwistor} sets all the OPE coefficients to zero, i.e.
\begin{align}
c^{(nh)}_{s_is_js_k}=0,\qquad\forall s_i,s_j,s_k.
\end{align}
Hence the other choice of supertwistors, i.e. $(W,\bar{\psi})$ for plus helicities and $(Z,\psi)$ for minus helicities (which correspond to non-homogeneous part) are inconsistent with supersymmetric Ward identities \eqref{QDualTwistor}.
\begin{align}\label{intnon}
\langle0| \mathbf{\hat{J}_{s_{1}}^+}(W_1,\bar{\psi}_1) \mathbf{\hat{J}_{s_{2}}^+}(W_2,\bar{\psi}_2) \mathbf{\hat{J}_{s_{3}}^+}(W_3,\bar{\psi}_3)|0\rangle_{nh}=0,  
\end{align}
where $s_i\in\mathbb{Z}_\geq0$. Thus, the supertwistor correlators of integer spin supercurrents do not admit the non-homogeneous contributions, which is perfectly in accord with \cite{Jain:2023idr}. 

It is noteworthy that the results until now are Grassmann even structures with a manifest dot product \eqref{supertwistorsdot1} of (dual) supertwistor variables (unlike \eqref{13+nhold}). This fact plays a crucial role in our understanding of super-correlators in supertwistor space, as illustrated in subsequent discussions. Let us now explore the cases involving conserved half integer spin supercurrents.

\subsection{Half integer spin supercurrent}
 Recall the conserved supercurrents \eqref{supertwistorcurrent1} and \eqref{superdualtwistorcurrent1}. While the two point functions are fine in either of the choices, the three point super-correlators involving the spin-half supercurrents take a Grassmann odd structure when expressed in variables $(Z,\psi)$ for plus helicities and $(W,\bar{\psi})$ for minus helicities. These super-correlators are thus incapable of rendering a Grassmann dot products. Instead, we shall deploy their conjugate variables to express super-correlators of the spin-half supercurrents (i.e. $(W,\bar{\psi})$ for plus helicities and $(Z,\psi)$ for minus helicities), which naturally results in Grassmann even structures. This leads to the non-homogeneous part for spin-half supercurrents, while the homogeneous part vanishes. Let us begin with the two point function of spin-half supercurrents.

\subsubsection{Two point functions}
The two point super-correlator for conserved supercurrents of half integer spins in plus helicity can be written using \eqref{supertwistorcurrent1} and \eqref{superdualtwistorcurrent1}. Although conformal invariance along with helicity identity constrains the component correlators, the action of the supersymmetric Ward identity \eqref{QDualTwistor} imposes constraints on the relative coefficients between the component correlators. The expression yet again takes the form of a binomial expansion, similar to the earlier cases. Thus, one can compactly express the two point function in (++) as,
\begin{align}\label{supertwistor2point++conjugate}
\langle0|\mathbf{\hat{J}}_{s}^+(W_1,\bar{\psi}_1) \mathbf{\hat{J}}_{s}^+(W_2,\bar{\psi}_2)|0\rangle
&= \frac{\textit{i}^{-2(s+\frac{1}{2})+2} \big(c_{s+\frac{1}{2}}^{(even)}+ic_{s+\frac{1}{2}}^{(odd)}\big)}{(W_1\cdot W_2+\bar{\psi}_{1}\bar{\psi}_{2})^{-2(s+\frac{1}{2})+2} }.
\end{align}
A similar derivation for (- -) helicity yields,
\begin{align}\label{supertwistor2point--conjugate}
\langle0|\mathbf{\hat{J}}_{s}^{-}(Z_1,\psi_1) \mathbf{\hat{J}}_{s}^-(Z_2,\psi_2)|0\rangle
&= \frac{(-\textit{i})^{-2(s+\frac{1}{2})+2} \big(c_{s+\frac{1}{2}}^{(even)}-ic_{s+\frac{1}{2}}^{(odd)}\big)}{(Z_1\cdot Z_2-{\psi}_{1}{\psi}_{2})^{-2(s+\frac{1}{2})+2} }.
\end{align}
Let us now move on to the three point functions.

\subsubsection{Three point functions}
We begin our analysis with the non-homogeneous part of three point functions for the conserved spin-half supercurrent.

\subsubsection*{Non-homogeneous correlators}
Consider the three point function for conserved supercurrents of half integer spins in plus helicity using \eqref{superdualtwistorcurrent1}. The component correlators are fixed by the conformal invariance along with helicity identity up to some coefficient, whereas the OPE coefficients are fixed upon demanding the invariance of super-correlator under supersymmetry \eqref{QDualTwistor}. One then simplifies the resultant expression using the integral representation of delta-function, which leads to,
\begin{align}\label{conjugate3+}
\langle0| \mathbf{\Tilde{J}_{s_{1}}^+}(W_1,\bar{\psi}_1) \mathbf{\Tilde{J}_{s_{2}}^+}(W_2,\bar{\psi}_2) \mathbf{\Tilde{J}_{s_{3}}^+}(W_3,\bar{\psi}_3)|0\rangle=&(-\textit{i})^{-s_1-s_2-s_3}c_{s_1s_2s_3}^{(nh)}\delta^{[-(s_1+s_2-s_3)]}(W_1\cdot W_2+\bar{\psi}_{1} \bar{\psi}_{2})\notag\\\delta^{[-(s_2+s_3-s_1)]}(W_2\cdot W_3+\bar{\psi}_{2} \bar{\psi}_{3})&\delta^{[-(s_3+s_1-s_2)]}(W_3\cdot W_1+\bar{\psi}_{3} \bar{\psi}_{1}).
\end{align}
We would like to emphasize on the point that the super-correlators for spin-half supercurrents are non-homogeneous, which is in agreement with \cite{Jain:2023idr}. Let us now solve for the homogeneous part of the super-correlator for a half integer spin supercurrent.

\subsubsection*{Homogeneous correlators}
A similar analysis for the three point function for conserved supercurrents of half integer spins in plus helicity using $(Z_i,\psi_i)$ variable \eqref{supertwistorcurrent1} should give the homogeneous part \cite{Bala:2025gmz}. However, the action of supersymmetry Ward-identity \eqref{QTwistor} sets the OPE coefficients of all the component correlators to zero, i.e.
\begin{align}
c^{(h)}_{s_is_js_k}=0,\qquad\forall s_i,s_j,s_k.
\end{align}
Thus, the homogeneous part of the three point function involving conserved spin-half supercurrents is inconsistent with supersymmetric Ward identities \eqref{QDualTwistor}, which is in agreement with \cite{Jain:2023idr}.
\begin{align}\label{halfhomo}
\langle0| \mathbf{\hat{J}_{s_{1}}^+}(Z_1,\psi_1) \mathbf{\hat{J}_{s_{2}}^+}(Z_2,\psi_2) \mathbf{\hat{J}_{s_{3}}^+}(Z_3,\psi_3)|0\rangle_{h}=0,  
\end{align}
where $s_i\in\frac{2\mathbb{Z}_{\geq 0}+1}{2}$. Let us now switch gears and explore cases with extended supersymmetries.

\section{Extended supersymmetries}\label{Extended}

Extended supersymmetries are a tricky avenue: while these theories enjoy a larger number of symmetries, they also have larger supercurrent multiplets.  For instance, while the $\mathcal{N}=1$ supercurrent goes from spin $s$ to $s+\frac{1}{2}$, the $\mathcal{N}$ extended supercurrent goes from spin $s$ to $s+\frac{\mathcal{N}}{2}$.\footnote{The superspace construction is valid only until $\mathcal{N}=4$.} This implies that the super-correlator now has a large number of component correlators, which amounts to a large number of structures to deal with. Moreover, the analysis becomes technically challenging and is progressively complicated for higher supersymmetries.

The extended supersymmetries, enjoy an $\mathfrak{so}(\mathcal{N})$ R-symmetry on top of the usual super-charge invariance. This allows one to view extended supersymmetries in the same light as $\mathcal{N}=1$ case with suitable changes. For an array of Grassmann variables $\psi_i^N,\;\bar{\psi}_i^N$, the supersymmetry generators take the following form,
\begin{align}\label{QExt} [Q_{a}^N(\lambda,\bar{\mu},\psi^N),\mathbf{\hat{J}_s^\pm}(\lambda, \bar{\mu},\psi^N)]   &= \frac{e^{-\frac{i\pi}{4}}}{2\sqrt{2}}\Big( \lambda_a \frac{\partial}{\partial \psi_N} + \psi^N \frac{\partial}{\partial \bar{\mu}^a}\Big) \mathbf{\hat{J}_s^\pm}(\lambda, \bar{\mu},\psi^N), \notag \\ [Q_{a}^N(\mu,\bar{\lambda},\bar{\psi}^N),\mathbf{\hat{J}_s^\pm}(\mu,\bar{\lambda},\bar{\psi}^N)]   &= \frac{e^{-\frac{i\pi}{4}}}{2\sqrt{2}}\Big(\bar{\lambda}_a \frac{\partial}{\partial \bar{\psi}_N} - \bar{\psi}^N \frac{\partial}{\partial \mu^a}\Big)\mathbf{\hat{J}_s^\pm}(\lambda, \bar{\mu},\bar{\psi}^N),
\end{align}
where $N=\{1,\cdots,\mathcal{N}\}$ denotes the vector index of the $\mathfrak{so}(\mathcal{N})$ R-symmetry. The conserved supercurrents too can be seen as an extension from the previous section.

The strategy for solving super-correlators in extended supersymmetries remains mostly the same as in the $\mathcal{N}=1$ case. The intricacies only lie in the manipulation of R-symmetry indices. Let us now solve for the two and three point super-correlators for $\mathcal{N}=2,3,4$ cases for integer spin supercurrents. 

\subsection{Integer spin supercurrent}
Consider the conserved supercurrents of integer spin. Unlike the $\mathcal{N}=1$ case, where the supercurrents were obtained by performing relevant supertwistor transformations starting from their momentum superspace counterparts; let us now adopt a different strategy: Recall that the supercurrents \eqref{supertwistorcurrent1} and \eqref{superdualtwistorcurrent1} carried a specific helicity\footnote{This has been checked explicitly for $\mathcal{N}$=2 where the supesupercurrent indeed has this feature. We assume this should work for higher $\mathcal{N}$ as well because the conserved supercurrents in three dimensions always have two helicities + and -.} i.e.
\begin{align}\label{helicitysupercurrent}
\mathbf{h}[\mathbf{\hat{J}^+_s}(Z,\psi_{A})]=+s \, \mathbf{\hat{J}^+_s}(Z,\psi_{A}) &\qquad\mathbf{h}[\mathbf{\hat{J}^-_s}(W,\bar{\psi}_{A})]=-s \, \mathbf{\hat{J}^-_s}(W,\bar{\psi}_{A}). 
\end{align}
Thus one can now leverage helicity arguments to construct supercurrents. \small
\begin{align}\label{supertwistorcurrentExt}
\mathbf{\hat{J}_s^+}(Z,\psi_{A})=& \left( \hat{J}_s^+ (Z) +e^{\frac{i\pi}{4}} \psi_{A_1}\hat{J}_{s+\frac{1}{2}}^{+A_1}(Z)+e^{\frac{i\pi}{2}} \psi_{A_1}\psi_{A_2}\hat{J}_{s+1}^{+A_1A_2}(Z)+\cdots+e^{\frac{i\mathcal{N}\pi}{4}} \psi_{A_1}...\psi_{A_\mathcal{N}}\hat{J}_{s+\frac{\mathcal{N}}{2}}^{+A_1...A_\mathcal{N}}(Z)\;\right),\notag \\
\mathbf{\hat{J}_s^-}(W,\bar{\psi}_{A})=& \left( \hat{J}_s^- \left(W\right) +e^{\frac{i\pi}{4}} \bar{\psi}_{A_1}\hat{J}_{s+\frac{1}{2}}^{-A_1}\left(W \right)+e^{\frac{i\pi}{2}}\bar{\psi}_{A_1}\bar{\psi}_{A_2}\hat{J}_{s+1}^{-A_1A_2}(W )+\cdots+e^{\frac{i\mathcal{N}\pi}{4}}\bar{\psi}_{A_1}...\bar{\psi}_{A_\mathcal{N}}\hat{J}_{s+\frac{\mathcal{N}}{2}}^{-A_1...A_\mathcal{N}}(W ) \right).
\end{align}
\normalsize Very importantly, note that we have properly accounted for factors of $e^{\frac{i\pi}{4}}$ appropriately to ensure the fact that the currents are real. See appendix \ref{app:realitysusy} for details.
Let us now start with solving for the two point super-correlators for integer spin supercurrents with extended supersymmetry.
\subsubsection{Two point functions}
The two point super-correlator for conserved supercurrents of integer spins in plus helicity can be written using \eqref{supertwistorcurrentExt}. One begins by writing the most general ansatz adhering to the supercurrent \eqref{supertwistorcurrentExt}. Then, we perform a tensor decomposition among the R-symmetry indices using various Levi-Civita identities\footnote{Some details of this step are presented in appendix \ref{app:tensor}.}. While the conformal invariance along with helicity identity constraints the component correlators, the action of supersymmetric Ward identity \eqref{QExt} imposes constraints on the relative coefficients between the component correlators. In principle the super-correlator will have terms proportional to parity even R-symmetry tensor structure $(\delta^{AB})$ and that proportional to parity odd\footnote{Here we are strictly talking about the parity structures of R-symmetry indices and not about the space-time coordinates.} R-symmetry tensor structure $(\epsilon^{A_1\cdots A_N})$. The supersymmetry generator \eqref{QExt} demands that the parity odd R-symmetry tensor structure must be zero. The expression yet again takes form of a binomial expansion upon rearranging, akin to the $\mathcal{N}=1$ two point function \eqref{12-}. Thus, one can compactly express the two point function as,
\begin{align}\label{22+}
\langle0|\mathbf{\hat{J}_{s}^+}\left(Z_1,\psi_{1M}\right) \mathbf{\hat{J}_{s}^+}(Z_2,\psi_{2N})|0\rangle =\frac{\textit{i}^{2s+2}\big(c_{s+\frac{1}{2}}^{(even)}+ic_{s+\frac{1}{2}}^{(odd)}\big)}{(Z_{1}\cdot Z_{2} - \delta^{MN}\psi_{1M}\psi_{2N})^{2(s+1)}}.
\end{align}
Notice that the denominator is yet again the supertwistor dot product \eqref{supertwistorsdotExt} now in accordance with $\mathfrak{so}(\mathcal{N})$ R-symmetry. Similarly for the minus helicity, we find that the two point function is,
\begin{align}\label{22-}
\langle0| \mathbf{\hat{J}_{s}^-}(W_1,\bar{\psi}_{1M}) \mathbf{\hat{J}_{s}^-}(W_2,\bar{\psi}_{2N})|0\rangle = \frac{(-\textit{i})^{2s+2}\big(c_{s+\frac{1}{2}}^{(even)}-ic_{s+\frac{1}{2}}^{(odd)}\big)}{(W_1\cdot W_2+\delta^{MN}\bar{\psi}_{1M}\bar{\psi}_{2N})^{2s+2}}. 
\end{align}
Let us now solve the three point functions for super-correlators of integer spin supercurrents with extended supersymmetry.

\subsubsection{Three point functions}
Consider the three point super-correlator for conserved supercurrents of integer spins in plus helicity, which can be written using \eqref{supertwistorcurrentExt}. Repeating the same analysis as above yields a compact result,
\begin{align}\label{Homogen}
\langle0| \mathbf{\hat{J}_{s_{1}}^+}(Z_1,\psi_{1I}) \mathbf{\hat{J}_{s_{2}}^+}(Z_2,\psi_{2J}) \mathbf{\hat{J}_{s_{3}}^+}(Z_3,\psi_{3K})|0\rangle=&(-\textit{i})^{s_1+s_2+s_3}c_{s_1s_2s_3}^{(h)}\delta^{[s_1+s_2-s_3]}(Z_{1}\cdot Z_{2} - \delta^{IJ}\psi_{1I}\psi_{2J})\notag\\
\delta^{[s_2+s_3-s_1]}(Z_{2}\cdot Z_{3} - \delta^{JK}\psi_{2J}\psi_{3K})&\delta^{[s_3+s_1-s_2]}(Z_{3}\cdot Z_{1} - \delta^{KI}\psi_{3K}\psi_{1I}),
\end{align}
where the dot products are yet again in accord with $\mathfrak{so}(\mathcal{N})$ R-symmetry. This is indeed the homogeneous part of the correlator \eqref{inthomo}, whereas the non-homogeneous part of the correlator \eqref{intnon} turns out to be zero for extended supersymmetries too. This is in agreement with \cite{Jain:2023idr}. Let us now explore cases of extended supersymmetries for half integer spin supercurrents.

\subsection{Half integer spin supercurrent}
For half integer spin supercurrents\footnote{The half integer spin supercurrents exist only for $\mathcal{N}=1,3$ supersymmetries.}, we shall deploy the conjugate variables which ensure a Grassmann even structure analogous to the $\mathcal{N}=1$ case. Using helicity arguments, the supercurrents are as follows,
\begin{align}\label{supertwistorcurrentExtConj}
\mathbf{\hat{J}_s^+}(W,\bar{\psi}_{A})=& \left(e^{\frac{i\mathcal{N}\pi}{4}} \bar{\psi}_{A_1}...\bar{\psi}_{A_\mathcal{N}}\hat{J}_{s}^{+A_1...A_\mathcal{N}}(W)+\cdots+\hat{J}_{s+\frac{\mathcal{N}}{2}}^+\left(W\right)\right),\notag \\
\mathbf{\hat{J}_s^-}(Z,\psi_{A})=& \left(e^{\frac{i\mathcal{N}\pi}{4}}  \psi_{A_1}...\psi_{A_\mathcal{N}}\hat{J}_{s}^{-A_1...A_\mathcal{N}}(Z)+\cdots+\hat{J}_{s+\frac{\mathcal{N}}{2}}^-\left(Z\right)\right),
\end{align}
where now the supercurrents of course satisfy,
\begin{align}\label{helicitysupercurrentConj}
\mathbf{h}[\mathbf{\hat{J}^+_s}]=+s \mathbf{\hat{J}^+_s},&\quad\mathbf{h}[\mathbf{\hat{J}^-_s}]=-s \mathbf{\hat{J}^-_s}. 
\end{align}
Let us start with two point functions of conserved supercurrents of half integer spin.
\subsubsection{Two point functions}
The two point super-correlator for conserved supercurrents of half integer spins in (++) helicity can be written using \eqref{supertwistorcurrentExtConj}. One repeats the same analysis as for earlier cases and express the two point function as,
\begin{align}\label{22+}
\langle0| \mathbf{\hat{J}_{s}^+}(W_1,\bar{\psi}_{1M}) \mathbf{\hat{J}_{s}^+}(W_2,\bar{\psi}_{2N})|0\rangle = \frac{\textit{i}^{-2(s+\frac{\mathcal{N}}{2})+2}\big(c_{s+\frac{\mathcal{N}}{2}}^{(even)}+ic_{s+\frac{\mathcal{N}}{2}}^{(odd)}\big)}{(W_1\cdot W_2+\delta^{MN}\bar{\psi}_{1M}\bar{\psi}_{2N})^{-2(s+\frac{\mathcal{N}}{2})+2}}. 
\end{align}
Similarly for the (- -) helicity the two point function is,
\begin{align}\label{22-}
\langle0| \mathbf{\hat{J}_{s}^-}\left(Z_1,\psi_{1M}\right) \mathbf{\hat{J}_{s}^-}(Z_2,\psi_{2N})|0\rangle =\frac{(-\textit{i})^{-2(s+\frac{\mathcal{N}}{2})+2}\big(c_{s+\frac{\mathcal{N}}{2}}^{(even)}-ic_{s+\frac{\mathcal{N}}{2}}^{(odd)}\big)}{(Z_{1}\cdot Z_{2} - \delta^{MN}\psi_{1M}\psi_{2N})^{-2(s+\frac{\mathcal{N}}{2})+2}}.
\end{align}
Let us now solve the three point functions for half integer spin supercurrents.
\subsubsection{Three point functions}
Using the supercurrents \eqref{supertwistorcurrentExtConj} in conjugate variables, solving for the superconformal Ward identities  and helicity identity yields the following result,
\begin{align}\label{Nonhomogen}
&\langle0| \mathbf{\hat{J}_{s_{1}}^+}(W_1,\psi_{1A}) \mathbf{\hat{J}_{s_{2}}^+}(W_2,\psi_{2B}) \mathbf{\hat{J}_{s_{3}}^+}(W_3,\psi_{3C})|0\rangle=(-\textit{i})^{-(s_1+s_2+s_3+\frac{3\mathcal{N}}{2})}c_{s_1s_2s_3}^{(nh)}\notag\\&\delta^{[-(s_1+s_2-s_3+\frac{\mathcal{N}}{2})]}(W_1\cdot W_2+\delta^{AB}\bar{\psi}_{1A} \bar{\psi}_{2B})
\delta^{[-(s_2+s_3-s_1+\frac{\mathcal{N}}{2})]}(W_2\cdot W_3+\delta^{BC}\bar{\psi}_{2B} \bar{\psi}_{3C})\notag\\&\delta^{[-(s_3+s_1-s_2+\frac{\mathcal{N}}{2})]}(W_3\cdot W_1+\delta^{CA}\bar{\psi}_{3C} \bar{\psi}_{1A}).
\end{align}
Thus the super-correlators for spin-half supercurrents are non-homogeneous for extended supersymmetries \cite{Jain:2023idr}.

So far we have observed that ensuring a manifest supertwistor structure not only captures superconformal invariance, but also the conservation of super-correlators and existence of correct structure (i.e. homogeneous for integer spin and non-homogeneous for half integer spin supesupercurrents.). Inspired by this fact, let us formulate a manifest supertwistor space in the following section.

\section{Solving in manifest supertwistor space}\label{sec:ManifestBootstrap}
Recall that in the previous sections, we used the fact that the CFT generators along with helicity operator constrain the component correlators while the supersymmetric generator $Q$ constrain their relative coefficients. This enables one to express the super-correlator compactly in terms of supertwistor dot products. However, there is a much more elegant way to achieve this. In non supersymmetric three dimensional CFTs \cite{Baumann:2024ttn,Bala:2025gmz}, conformal invariance in twistor space amounts to $Sp(4;\mathbb{R})$ invariance, which fixes the solutions as functions of (dual) twistor dot products.\footnote{This statement is modulo the existence of four dimensional delta functions like $\delta^{(4)}(\sum_ic_iZ_i^A)$, which too are $Sp(4;\mathbb{R})$ invariant.} One then needs a helicity operator that fixes the functional form of the two and three point functions of conserved currents completely.

Similarly for the supersymmetric scenario, one can use the supergroup $OSp(\mathcal{N}|4;\mathbb{R})$ to solve the super-correlator in a manifest manner. We present the superconformal generator \eqref{supertwistorgeneratorZ} in supertwistor space here again for convenience,
\begin{align}\label{supertwistorgeneratorZnew}
\mathcal{T}^{\mathcal{AB}}(\mathcal{Z}) \equiv\mathcal{Z}^{\mathcal{(A}} \frac{\partial}{\partial\mathcal{Z}_{\mathcal{\mathcal{B}]}}}= \mathcal{Z}^{\mathcal{A}} \frac{\partial}{\partial\mathcal{Z}_{\mathcal{B}}} + (-1)^{\eta_{\mathcal{A}}\eta_{\mathcal{B}}} \mathcal{Z}^{\mathcal{B}} \frac{\partial}{\partial\mathcal{Z}_{\mathcal{A}}}.
\end{align}
In the supertwistor variables, the helicity counting identities are of paramount importance. In the manifest supertwistor variables, the helicity identities \eqref{HTwistor} and \eqref{HDual} translate to the equations \eqref{superhelicity1} and \eqref{superhelicity2}.
The on-shell superamplitudes for $\mathcal{N}=4$ SYM in four dimensions enjoy superconformal invariance where the helicity operator is a part of the on-shell superconformal algebra \cite{Elvang:2013cua}. While it is not the same in our off-shell superspace formalism, the helicity operator does commute with all the superconformal generators for SCFT$_3$. As observed in \cite{Bala:2025gmz}, it plays a crucial role in fixing the functional form of component correlators. Its important role continues in the supersymmetric setting as well.

Equipped with the superconformal generators \eqref{supertwistorgeneratorZnew} and the helicity operators \eqref{superhelicity1}, \eqref{superhelicity2} in our arsenal, let us now systematically reformulate the supertwistor space analysis in these manifest variables. Adhering to the conventions of section \ref{N1}, we use $\mathcal{Z/W}$ for +/- helicity for integer spin supercurrents and vice-versa for half integer spin supercurrents. Let us start with the cases involving integer spin supercurrents.

\subsection{Integer spin supercurrent}
We begin by solving super-correlators involving conserved supercurrents with integer spin. The manifest supertwistor analysis captures the correct structure for three point functions i.e. the homogeneous part. Let us begin with solving two point function in these manifest variables.
\subsubsection{Two point super-correlators}
Consider the two point function with positive-helicity in the $\mathcal{Z}$ representation.\footnote{We have observed in earlier sections that $\mathcal{Z}$ for plus helicity and $\mathcal{W}$ for minus helicity are the natural choices for integer spin supercurrent.} The action of superconformal Ward identities \eqref{manifesttwistorconformalWard} on super-correlators are given as,
\begin{align}\label{ManifestSCFTWI}
\Bigg(\mathcal{Z}_{1}^{(\mathcal{A}}\frac{\partial}{\partial \mathcal{Z}_{1\mathcal{B}]}}+ \mathcal{Z}_{2}^{(\mathcal{A}}\frac{\partial}{\partial \mathcal{Z}_{2\mathcal{B}]}}\Bigg)\langle 0|\hat{\mathbf{J}}_s^{+}(\mathcal{Z}_1)\hat{\mathbf{J}}_s^{+}(\mathcal{Z}_2)|0\rangle=0.
\end{align}
Just like in the case of conformal symmetry \cite{Bala:2025gmz}, the solution to the above differential equation simply fixes the argument,\footnote{We present the details of solving this differential equation in appendix \ref{app:WI}.}
\begin{align}
    \langle 0|\hat{\mathbf{J}}_s^{+}(\mathcal{Z}_1)\hat{\mathbf{J}}_s^{+}(\mathcal{Z}_2)|0\rangle = F(\mathcal{Z}_1 \cdot \mathcal{Z}_2).
\end{align}
The helicity identity \eqref{superhelicity1} then constrains the functional form with the following equations,
\begin{align}\label{superhelicityaction2point}
\mathcal{Z}_{1\mathcal{A}}\frac{\partial}{\partial \mathcal{Z}_{1\mathcal{A}}}F(\mathcal{Z}_1\cdot \mathcal{Z}_2) = -2(s+1) F(\mathcal{Z}_1\cdot \mathcal{Z}_2),\quad\mathcal{Z}_{2\mathcal{A}}\frac{\partial}{\partial \mathcal{Z}_{2\mathcal{A}}} F(\mathcal{Z}_1\cdot \mathcal{Z}_2) = -2(s+1) F(\mathcal{Z}_1\cdot\mathcal{Z}_2).
\end{align}
The helicity differential equation can be recast into an Euler equation which has two distinct types of solutions which fall into the class of generalized functions \cite{kanwal1998generalized}. While the \textit{polynomial function} is an obvious candidate, there is a \textit{distributional} solution as well. The detailed analysis of these solutions are presented for twistor space in \cite{Bala:2025gmz}. The eigenvalue equations \eqref{superhelicityaction2point} imply the following two solutions\footnote{We present the details of solving the helicity differential equation in appendix \ref{app:helicity}.},
\begin{align}\label{supertwistortwopointpp}
    \langle 0|\hat{\mathbf{J}}_s^{+}(\mathcal{Z}_1)\hat{\mathbf{J}}_s^{+}(\mathcal{Z}_2)|0\rangle\supset \bigg\{\frac{1}{(\mathcal{Z}_1\cdot \mathcal{Z}_2)^{2s+2}},\;\delta^{[2s+1]}(\mathcal{Z}_1\cdot\mathcal{Z}_2)\bigg\}.
\end{align}
Similarly, the candidates for two point function for (- -) in $\mathcal{W}$ representation are,
\begin{align}\label{dualsupertwistortwopointmm}
\langle 0|\hat{\mathbf{J}}_s^{-}(\mathcal{W}_1)\hat{\mathbf{J}}_s^{-}(\mathcal{W}_2)|0\rangle\supset& \bigg\{\frac{1}{(\mathcal{W}_1\cdot \mathcal{W}_2)^{2s+2}},\;\delta^{[2s+1]}(\mathcal{W}_1\cdot\mathcal{W}_2)\bigg\}.
\end{align}
We shall only consider the solution with polynomial form as it is the correct candidate for two point functions \cite{Bala:2025gmz,Baumann:2024ttn}. The delta-function solution here leads to incorrect results when converted to momentum-space spinor-helicity results \cite{Bala:2025gmz}.
\begin{align}\label{ManifestTwoPoint}
\langle 0|\hat{\mathbf{J}}_s^{+}(\mathcal{Z}_1)\hat{\mathbf{J}}_s^{+}(\mathcal{Z}_2)|0\rangle&=\frac{\textit{i}^{2s+2}(c_{s}^{(even)}+ic_{s}^{(odd)})}{(\mathcal{Z}_1\cdot \mathcal{Z}_2)^{2s+2}},\notag\\
\langle 0|\hat{\mathbf{J}}_s^{-}(\mathcal{W}_1)\hat{\mathbf{J}}_s^{-}(\mathcal{W}_2)|0\rangle&=\frac{(-\textit{i})^{2s+2} (c_{s}^{(even)}-ic_{s}^{(odd)})}{(\mathcal{W}_1\cdot \mathcal{W}_2)^{2s+2}}.
\end{align}
Let us now move on to solve the three point functions.
\subsubsection{Three point super-correlators}
One can perform a similar analysis for three point functions as well. The manifest supertwistor analysis captures the correct structure for three point functions i.e. the homogeneous part \eqref{Homogen}. Let us start with the $(+++)$ configuration with all currents in the $\mathcal{Z}$ representation. The superconformal invariance \eqref{supertwistorgeneratorZnew} for the super-correlators implies,
\begin{align}\label{3pointsuperconformalwardI}
 &\Bigg(\mathcal{Z}_{1}^{(\mathcal{A}}\frac{\partial}{\partial \mathcal{Z}_{1\mathcal{B}]}}+ \mathcal{Z}_{2}^{(\mathcal{A}}\frac{\partial}{\partial \mathcal{Z}_{2\mathcal{B}]}}+ \mathcal{Z}_{3}^{(\mathcal{A}}\frac{\partial}{\partial \mathcal{Z}_{3\mathcal{B}]}}\Bigg)\langle 0|\hat{\mathbf{J}}_{s_1}^{+}(\mathcal{Z}_1)\hat{\mathbf{J}}_{s_2}^{+}(\mathcal{Z}_2)\hat{\mathbf{J}}_{s_3}^{+}(\mathcal{Z}_3)|0\rangle = 0.
\end{align}
The solutions to above equation are simply functions with fixed arguments,
\begin{align}\label{3pointsupertwistorsol}
    &\langle 0| \hat{\mathbf{J}}^{+}_{s_1}(\mathcal{Z}_1)\hat{\mathbf{J}}^{+}_{s_2}(\mathcal{Z}_2)\hat{\mathbf{J}}^{+}_{s_3}(\mathcal{Z}_3)|0\rangle=F(\mathcal{Z}_1\cdot \mathcal{Z}_2,\mathcal{Z}_2\cdot \mathcal{Z}_3,\mathcal{Z}_3\cdot \mathcal{Z}_1).
\end{align}
The helicity identities \eqref{superhelicity1} constrains the functional form as,
\begin{align}\label{superhelicity3point}
\mathcal{Z}_{1\mathcal{A}}\frac{\partial}{\partial \mathcal{Z}_{1\mathcal{A}}}F=-2(s_1+1)F,~\mathcal{Z}_{2\mathcal{A}}\frac{\partial}{\partial \mathcal{Z}_{2\mathcal{A}}}F=-2(s_2+1)F,~\mathcal{Z}_{3\mathcal{A}}\frac{\partial}{\partial \mathcal{Z}_{3\mathcal{A}}}F=-2(s_3+1)F.
\end{align}
These differential equations yields two solutions: polynomial function or delta-function (distribution)\footnote{We present the analysis for cases where spin-triangle inequality is satisfied i.e. $s_i+s_j\leq s_k\;\forall\; i,j,k$; however it shall hold for other cases as well where one can generalize the delta-function representation to negative derivatives. Refer \cite{Bala:2025gmz} for further details in twistor space context.},
\begin{align}\label{pppsupertwistor}
    \langle 0|\hat{\mathbf{J}}^{+}_{s_1}(\mathcal{Z}_1)\hat{\mathbf{J}}^{+}_{s_2}(\mathcal{Z}_2)\hat{\mathbf{J}}^{+}_{s_3}(\mathcal{Z}_3)|0\rangle \supset \bigg\{& \delta^{[s_1+s_2-s_3]}(\mathcal{Z}_1\cdot \mathcal{Z}_2)\delta^{[s_2+s_3-s_1]}(\mathcal{Z}_2\cdot \mathcal{Z}_3)\delta^{[s_3+s_1-s_2]}(\mathcal{Z}_3\cdot \mathcal{Z}_1),\notag\\
    & \frac{1}{(\mathcal{Z}_1\cdot\mathcal{Z}_2)^{s_1+s_2-s_3+1}} \frac{1}{(\mathcal{Z}_2\cdot\mathcal{Z}_3)^{s_2+s_3-s_1+1}} \frac{1}{(\mathcal{Z}_3\cdot\mathcal{Z}_1)^{s_3+s_1-s_2+1}}\bigg\}.
\end{align}
It has been noticed in \cite{Bala:2025gmz,Baumann:2024ttn} that the solution with delta-function form as it is the correct candidate for three point functions. The polynomial solution here leads to incorrect results when converted to momentum-space spinor-helicity results \cite{Bala:2025gmz}.
\begin{align}\label{ManifestThreePoint}
 \langle 0|\hat{\mathbf{J}}^{+}_{s_1}(\mathcal{Z}_1)\hat{\mathbf{J}}^{+}_{s_2}(\mathcal{Z}_2)\hat{\mathbf{J}}^{+}_{s_3}(\mathcal{Z}_3)|0\rangle=&\;(-\textit{i})^{s_1+s_2+s_3}(c_{s_1s_2s_3}^{(h)}+ic_{s_1s_2s_3}^{(odd)})\notag\\&\delta^{[s_1+s_2-s_3]}(\mathcal{Z}_1\cdot \mathcal{Z}_2)\delta^{[s_2+s_3-s_1]}(\mathcal{Z}_2\cdot \mathcal{Z}_3)\delta^{[s_3+s_1-s_2]}(\mathcal{Z}_3\cdot \mathcal{Z}_1).
\end{align}
Note that the result corresponds to homogeneous part of the super-correlator while the non-homogeneous part vanishes. Let us now briefly summarize the three point functions for all the helicity-configurations.
\subsubsection*{Other helicities}
The three point function in all the helicities can be summarized succinctly as follows,
\begin{numcases}{\langle 0| \mathbf{\hat{J}}_{s_1}^{h_1}(\mathcal{T}_1)\mathbf{\hat{J}}_{s_2}^{h_2}(\mathcal{T}_2)\mathbf{\hat{J}}_{s_3}^{h_3}(\mathcal{T}_3)|0\rangle=}
(c_{s_1s_2s_3}^{(h)}-i c_{s_1s_2s_3}^{(odd)})(\textit{i})^{s_1+s_2+s_3}\notag \\ \delta^{[s_1+s_2-s_3]}(\mathcal{T}_1\cdot \mathcal{T}_2)\delta^{[s_2+s_3-s_1]}(\mathcal{T}_2\cdot \mathcal{T}_3)\delta^{[s_3+s_1-s_2]}(\mathcal{T}_3\cdot \mathcal{T}_1), \nonumber \\
\text{when sign}(h_1)+\text{sign}(h_2)+\text{sign}(h_3)<0, \nonumber\\
\nonumber\\
(c_{s_1s_2s_3}^{(h)}+i c_{s_1s_2s_3}^{(odd)})(-\textit{i})^{s_1+s_2+s_3}\notag \\\delta^{[s_1+s_2-s_3]}(\mathcal{T}_1\cdot \mathcal{T}_2)\delta^{[s_2+s_3-s_1]}(\mathcal{T}_2\cdot \mathcal{T}_3) \delta^{[s_3+s_1-s_2]}(\mathcal{T}_3\cdot \mathcal{T}_1),\nonumber\\
\text{when sign}(h_1)+\text{sign}(h_2)+\text{sign}(h_3)>0.
\end{numcases}
where $\mathcal{T}_i=\mathcal{Z}_i~\text{if}~h_i=+ ~\text{and}~ \mathcal{T}_i=\mathcal{W}_i ~\text{if}~ h_i=-$. Note that $c_{s_1s_2s_3}^{(h)} \And c_{s_1s_2s_3}^{(odd)}$ are coefficients corresponding to parity even homogeneous and parity odd parts respectively. The factors of $-i$ and $+i$ are necessary in order to maintain consistency with the spinor helicity results. Moreover, the parity odd part exists only when the spin-triangle inequality is satisfied i.e. $s_i+s_j\geq s_k\;\forall\;i,j,k$. Refer \cite{Bala:2025gmz} for further details. Let us now investigate the case of half integer spin supercurrents.

\subsection{Half integer spin supercurrent}
We will now solve super-correlators involving conserved supercurrents of half integer spins. The manifest supertwistor analysis captures the correct structure for three point functions i.e. the non-homogeneous part \eqref{Nonhomogen}. Let us begin with solving two point function in these manifest variables.
\subsubsection{Two point super-correlators}
Consider the two point function of positive-helicity in $\mathcal{W}$ representation. The action of superconformal Ward identities \eqref{manifesttwistorconformalWard} on super-correlators are given as,
\begin{align}
\Bigg(\mathcal{W}_{1}^{(\mathcal{A}}\frac{\partial}{\partial \mathcal{W}_{1\mathcal{B}]}}+ \mathcal{W}_{2}^{(\mathcal{A}}\frac{\partial}{\partial \mathcal{W}_{2\mathcal{B}]}}\Bigg)\langle 0|\hat{\mathbf{J}}_s^{+}(\mathcal{W}_1)\hat{\mathbf{J}}_s^{+}(\mathcal{W}_2)|0\rangle=0.
\end{align}
The solution to the superconformal Ward identity fixes the argument,
\begin{align}
    \langle 0|\hat{\mathbf{J}}_s^{+}(\mathcal{W}_1)\hat{\mathbf{J}}_s^{+}(\mathcal{W}_2)|0\rangle = F(\mathcal{W}_1 \cdot \mathcal{W}_2).
\end{align}
The helicity identity \eqref{superhelicity2} then constrains the functional form with the following equations\footnote{We deploy the conjugate variables i.e. $\mathcal{W}$ for plus helicity and $\mathcal{Z}$ for minus helicity for half integer spin supercurrents. Please note that the helicity eigenvalue for a supercurrent in the conjugate variable is $s+\frac{\mathcal{N}}{2}$ as observed in the previous analysis. This can be understood by a super-Fourier transform from $\mathcal{Z}$ to $\mathcal{W}$. Refer \ref{app:Fourier} for a detailed analysis.},
\begin{align}\label{superhelicityaction2pointodd}
\mathcal{W}_{1\mathcal{A}}\frac{\partial}{\partial \mathcal{W}_{1\mathcal{A}}}F(\mathcal{W}_1\cdot \mathcal{W}_2) = \Big(2\Big(s+\frac{\mathcal{N}}{2}\Big)-2\Big) F(\mathcal{W}_1\cdot \mathcal{W}_2),\notag\\
\mathcal{W}_{2\mathcal{A}}\frac{\partial}{\partial \mathcal{W}_{2\mathcal{A}}} F(\mathcal{W}_1\cdot \mathcal{W}_2) = \Big(2\Big(s+\frac{\mathcal{N}}{2}\Big)-2\Big) F(\mathcal{W}_1\cdot\mathcal{W}_2).
\end{align}
Solving the above equation fixes the two point functions in the following manner\footnote{The delta-function solution here leads to incorrect results for two point functions when converted to momentum-space spinor-helicity results.},
\begin{align}\label{ManifestTwoPointPP}
\langle 0|\hat{\mathbf{J}}_s^{+}(\mathcal{W}_1)\hat{\mathbf{J}}_s^{+}(\mathcal{W}_2)|0\rangle&=\frac{\textit{i}^{-2(s+\frac{\mathcal{N}}{2})+2}\big(c_{s+\frac{\mathcal{N}}{2}}^{even}+ic_{s+\frac{\mathcal{N}}{2}}^{odd}\big)}{(\mathcal{W}_1\cdot \mathcal{W}_2)^{-2\big(s+\frac{\mathcal{N}}{2}\big)+2}}.
\end{align}
The two point function in minus helicity similarly yields,
\begin{align}\label{ManifestTwoPointMM}
\langle 0|\hat{\mathbf{J}}_s^{-}(\mathcal{Z}_1)\hat{\mathbf{J}}_s^{-}(\mathcal{Z}_2)|0\rangle&=\frac{(-\textit{i})^{-2(s+\frac{\mathcal{N}}{2})+2}\big(c_{s+\frac{\mathcal{N}}{2}}^{even}-ic_{s+\frac{\mathcal{N}}{2}}^{odd}\big)}{(\mathcal{Z}_1\cdot \mathcal{Z}_2)^{-2\big(s+\frac{\mathcal{N}}{2}\big)+2}}.
\end{align}
Let us now solve the three point functions in the manifest variables.
\subsubsection{Three point super-correlators}
Let us start with the $(+++)$ configuration with all currents in the $\mathcal{W}$ representation. The superconformal invariance \eqref{manifesttwistorconformalWard} for the super-correlators  implies,
\begin{align}\label{3pointsuperconformalwardIodd}
 &\Bigg(\mathcal{W}_{1}^{(\mathcal{A}}\frac{\partial}{\partial \mathcal{W}_1^{\mathcal{B}]}}+ \mathcal{W}_{2}^{(\mathcal{A}}\frac{\partial}{\partial \mathcal{W}_{2\mathcal{B}]}}+ \mathcal{W}_{3}^{(\mathcal{A}}\frac{\partial}{\partial \mathcal{W}_{3\mathcal{B}]}}\Bigg)\langle 0|\hat{\mathbf{J}}_{s_1}^{+}(\mathcal{W}_1)\hat{\mathbf{J}}_{s_2}^{+}(\mathcal{W}_2)\hat{\mathbf{J}}_{s_3}^{+}(\mathcal{W}_3)|0\rangle = 0.
\end{align}
The solutions to above equations are as follows,
\begin{align}\label{3pointsupertwistorsolodd}
    &\langle 0| \hat{\mathbf{J}}^{+}_{s_1}(\mathcal{W}_1)\hat{\mathbf{J}}^{+}_{s_2}(\mathcal{W}_2)\hat{\mathbf{J}}^{+}_{s_3}(\mathcal{W}_3)|0\rangle=F(\mathcal{W}_1\cdot \mathcal{W}_2,\mathcal{W}_2\cdot \mathcal{W}_3,\mathcal{W}_3\cdot \mathcal{W}_1).
\end{align}
The helicity identities \eqref{superhelicity2} then constrains the functional form,
\begin{align}\label{superhelicity3pointodd}
\mathcal{W}_{1\mathcal{A}}\frac{\partial}{\partial \mathcal{W}_{1\mathcal{A}}}F=2\Big(s_1+\frac{\mathcal{N}}{2}-1\Big)F,\qquad&\mathcal{W}_{2\mathcal{A}}\frac{\partial}{\partial \mathcal{W}_{2\mathcal{A}}}F=2\Big(s_2+\frac{\mathcal{N}}{2}-1\Big)F, \notag \\ \mathcal{W}_{3\mathcal{A}}\frac{\partial}{\partial \mathcal{W}_{3\mathcal{A}}}F=&2\Big(s_3+\frac{\mathcal{N}}{2}-1\Big)F.
\end{align}
Solving the above equations, the three point functions yet again take a compact form\footnote{The polynomial solution here leads to incorrect results for three point functions when converted to momentum-space spinor-helicity results.},
\begin{align}\label{ManifestThreePointPPP}
 \langle 0|\hat{\mathbf{J}}^{+}_{s_1}(\mathcal{W}_1)\hat{\mathbf{J}}^{+}_{s_2}(\mathcal{W}_2)\hat{\mathbf{J}}^{+}_{s_3}(\mathcal{W}_3)|0\rangle=\;&(-\textit{i})^{-(s_1+s_2+s_3+\frac{3\mathcal{N}}{2})}c_{s_1s_2s_3}^{(nh)}\delta^{[-(s_1+s_2-s_3+\frac{\mathcal{N}}{2})]}(\mathcal{W}_1\cdot \mathcal{W}_2)\notag\\&\delta^{[-(s_2+s_3-s_1+\frac{\mathcal{N}}{2})]}(\mathcal{W}_2\cdot \mathcal{W}_3)\delta^{[-(s_3+s_1-s_2+\frac{\mathcal{N}}{2})]}(\mathcal{W}_3\cdot \mathcal{W}_1).
\end{align}
Note that the result corresponds to non-homogeneous part of the super-correlator while the homogeneous part vanishes.

\subsubsection*{Other helicities}
    Performing a similar analysis, we get the following expression for other helicity configurations,
        \begin{numcases}
            {\langle 0| \mathbf{\hat{J}_{s_1}^{h_1}}(\mathcal{I}_1)\mathbf{\hat{J}_{s_2}^{h_2}}(\mathcal{I}_2)\mathbf{\hat{J}_{s_3}^{h_3}}(\mathcal{I}_3)|0\rangle=}
            c_{s_1s_2s_3}^{(nh)}(\textit{i})^{-(s_1+s_2+s_3+\frac{\mathcal{N}}{2})}\delta^{[-(s_1+s_2-s_3+\frac{\mathcal{N}}{2})]}(\mathcal{I}_1\cdot \mathcal{I}_2) \notag \\\delta^{[-(s_2+s_3-s_1+\frac{\mathcal{N}}{2})]}(\mathcal{I}_2\cdot \mathcal{I}_3)\delta^{[-(s_3+s_1-s_2+\frac{\mathcal{N}}{2})]}(\mathcal{I}_3\cdot \mathcal{I}_1), \nonumber \\
            \text{when sign}(h_1)+\text{sign}(h_2)+\text{sign}(h_3)<0, \nonumber\\
            \nonumber\\
            c_{s_1s_2s_3}^{(nh)}(-\textit{i})^{-(s_1+s_2+s_3+\frac{\mathcal{N}}{2})}\delta^{[-(s_1+s_2-s_3+\frac{\mathcal{N}}{2})]}(\mathcal{I}_1\cdot \mathcal{I}_2)\notag \\ \delta^{[-(s_2+s_3-s_1+\frac{\mathcal{N}}{2})]}(\mathcal{I}_2\cdot \mathcal{I}_3)\delta^{[-(s_3+s_1-s_2+\frac{\mathcal{N}}{2})]}(\mathcal{I}_3\cdot \mathcal{I}_1),\nonumber\\
            \text{when sign}(h_1)+\text{sign}(h_2)+\text{sign}(h_3)>0.
        \end{numcases}
    where $\mathcal{I}_i=\mathcal{W}_i~\text{if}~\mathbf{h}_i=+ ~\text{and}~ \mathcal{I}_i=\mathcal{Z}_i ~\text{if}~ \mathbf{h}_i=-$. Note that $c_{s_1s_2s_3}^{(nh)}$ is the coefficient that corresponds to the non-homogeneous part.
    One can perform a similar analysis for mixed super-correlators involving both integer and half integer spin supercurrent. We postpone such discussions to future work.

\section{Discussion and future direction}\label{Discussion}
    In this paper, we have introduced a manifest off-shell supertwistor space framework for three dimensional $\mathcal{N}=1, 2, 3, 4$ superconformal field theories, harnessing the corresponding $OSp(\mathcal{N}|4;\mathbb{R})$ supergroup. We solved two and three point correlators of (half) integer spin conserved supercurrents by using the generators of the supergroup, which explicitly reflect supersymmetry, R-symmetry, conformal invariance, and conservation laws. In addition to the superconformal generators, we find that superhelicity operators are essential for determining the functional form of these correlators. These equations take the form of first order graded Euler equations which, besides the usual polynomial solutions, also admit weak solutions in the form of distributions. Our results are a simple and elegant generalization of non-supersymmetric results. We also connect the component correlators to super-correlators, which explicitly solve the superconformal Ward identities. In this paper, we have focused on solutions that depend on supertwistor dot products. However, as alluded in footnote \ref{footnote}, there also exist solutions of the form $\delta^{(4|\mathcal{N})}(\sum_i c_i\mathcal{Z}_i^\mathcal{A})$, which shall form part of the subject of an upcoming work \cite{Bala:2025new}. There are several interesting avenues for further exploration, some of which are presented below.

\subsection*{Supersymmetric Penrose transform}
It is a well established idea that the twistor space correlation functions are closely related to position-space Wightman functions through a Penrose transform, which acts as a bridge between these two formulations. It thus allows to study Wightman functions in twistor space, where they take an extremely simple form. Establishing a similar relationship between supertwistor space and position superspace would be intriguing because it could provide insights into supersymmetric field theories, potentially simplifying the computation of correlation functions in superspace. Schematically, it should take the form which is the super generalization of the Penrose transformation of \cite{Baumann:2024ttn},  
\begin{align}
    \langle 0|J_{s_1}\cdots J_{s_n}|0\rangle=\int D\mathcal{Z}_1\cdots D\mathcal{Z}_n~(\mathcal{Z}_1\cdot \Upsilon_1^*)^{2s_1}\cdots (\mathcal{Z}_n\cdot \Upsilon_n^*)^{2s_n}F(\mathcal{Z}_1,\cdots,\mathcal{Z}_n),
\end{align}
We hope to come back to this problem in the near future.
\subsection*{Super BCFW relations and Yangian symmetry}
BCFW recursion relations are a powerful tool for calculating scattering amplitudes in gauge theories. It would be fascinating to establish a similar technique for superconformal correlators, which shall allow to recursively obtain higher-point functions in terms of lower-point ones in a systematic and efficient way. In that vein, implementing dual superconformal symmetry and Yangian symmetry (which have led to remarkable discoveries for superamplitudes, see \cite{Arkani-Hamed:2010zjl,Adamo:2011pv,Arkani-Hamed:2012zlh,Arkani-Hamed:2013jha}) also presents an interesting direction to pursue for CFT super-correlators. 

\subsection*{Supersymmetric Chern-Simons matter theories}
Supersymmetric Chern-Simons matter theories are an interesting class of three dimensional field theories. For example, in \cite{Inbasekar:2017ieo,Inbasekar:2017sqp,Inbasekar:2020hla}, the authors show that all tree-level scattering amplitudes can be bootstrapped using the BCFW recursion relation in $\mathcal{N}=2$ supersymmetric Chern-Simons theory. It would be interesting to see if the supertwistors that we have employed in this paper can aid in the study of such theories.

Developing a similar construction as we have done for higher supersymmetry ($\mathcal{N}>4$) is also an important task. For example, this can be used to tackle problems in the $\mathcal{N}=6$ ABJ(M) theory. It would be nice to emulate the successes in the study of scattering amplitudes in this theory with respect to correlation functions.  We hope supertwistors can serve as a tool to study these off-shell observables.

\subsection*{Non-conserved currents in supertwistor space}
In four dimensions field theories with supersymmetry, BPS states play a key role in analysis of theories which preserve some amount of supersymmetry. It would be important to develop a similar extension to algebra for the three dimensional case. This can possibly aid in the analyses involving non-conserved currents in an off-shell manner by implementing an analogue of the BPS condition.

\acknowledgments
We acknowledge our debt to the people of India for their constant support of research in basic sciences. We thank Mohd.Ali, Nipun Bhave, Amin Nizami, and Saurabh Pant for useful discussions. AB acknowledges a UGC-NET fellowship.

\appendix
\section{Notations and conventions}\label{app:note}
The contractions of supertwistors $\mathcal{Z}^\mathcal{A}$ and $\mathcal{W}_\mathcal{A}$ are done using the invariant tensor $\Omega_{AB}$ of the $OSp(\mathcal{N}|4;\mathbb{R})$ super group given by,
\begin{align}
\Omega_{\mathcal{A}\mathcal{B}}\equiv\Omega^{\mathcal{A}\mathcal{B}}=\begin{pmatrix}
        0&\delta_a^{b}&0\\
        -\delta^b_{a} & 0&0\\
        0&0&\mathbb{I}_{\mathcal{N}\cross \mathcal{N}}
    \end{pmatrix}.
\end{align}
The contraction of the invariant tensor is given by,
\begin{align}
\Omega_{\mathcal{A}\mathcal{B}}\Omega^{\mathcal{A}\mathcal{C}}= \delta^{\mathcal{C}}_{\mathcal{B}}.
\end{align} 
The infinitesimal supersymmetric generator \eqref{supertwistorgeneratorZ} $\mathcal{T}_{\mathcal{AB}}$ follows,
\begin{align}
    \mathcal{T}_{\mathcal{A}\mathcal{B}}= (-1)^{\eta_{\mathcal{A}}\eta_{\mathcal{B}}} \mathcal{T}_{\mathcal{B}\mathcal{A}}.
\end{align}
The objects ($\mathcal{X}, \mathcal{Y}$) that transform covariantly under the $OSp(\mathcal{N}|4;\mathbb{R})$ supergroup follow certain properties. For instance, the exchange of these vectors follows, 
\begin{align}
\mathcal{X}_{\mathcal{A}} \mathcal{Y}_{\mathcal{B}} = (-1)^{\eta_{\mathcal{A}}\eta_{\mathcal{B}}} \mathcal{Y}_{\mathcal{B}} \mathcal{X}_{\mathcal{A}}.
\end{align}
We define the graded commutator in the indices as,
\begin{align}
     \mathcal{X}_{(A} \mathcal{Y}_{B]} \equiv \mathcal{X}_{\mathcal{A}} \mathcal{Y}_{\mathcal{B}} + (-1)^{\eta_{\mathcal{A}}\eta_{\mathcal{B}}} \mathcal{X}_{\mathcal{B}} \mathcal{Y}_{\mathcal{A}}, \qquad \mathcal{X}_{[A} \mathcal{Y}_{B)} \equiv \mathcal{X}_{\mathcal{A}} \mathcal{Y}_{\mathcal{B}} - (-1)^{\eta_{\mathcal{A}}\eta_{\mathcal{B}}} \mathcal{X}_{\mathcal{B}} \mathcal{Y}_{\mathcal{A}}.
\end{align}
 The raising/lowering of (anti-)fundamental representation index of $OSp(\mathcal{N}|4;\mathbb{R}$) is done by,
\begin{align}
      \Omega_{\mathcal{A}\mathcal{B}} \mathcal{X}^{\mathcal{A}} = \mathcal{X}_\mathcal{B}, \qquad \Omega^{\mathcal{A}\mathcal{B}} \mathcal{X}_{\mathcal{B}} = \mathcal{X}^{\mathcal{A}} \notag  .  
\end{align}
The super twistor dot product is defined by\footnote{There is an exception for the supertwistor dot product for $\mathcal{Z}$ i.e. $\mathcal{Z}_{i}\cdot\mathcal{Z}_{j} \equiv -\mathcal{Z}_{i\mathcal{A}}\mathcal{Z}_{j}^{\mathcal{A}}$. We define it with an extra minus sign for convenience.}
\begin{align}
      \mathcal{X} \cdot \mathcal{Y} = \mathcal{X}_{\mathcal{A}} \mathcal{Y}^{\mathcal{A}} =\mathcal{X}^{\mathcal{B}} \Omega_{\mathcal{B}\mathcal{A}} \mathcal{Y}^{\mathcal{A}} = \mathcal{X}_{\mathcal{A}} \Omega^{\mathcal{A}\mathcal{B}}\mathcal{Y}_{\mathcal{B}} \notag.
\end{align}

\section{Superconformal algebra in supertwistor variables}\label{app:sup}
In the following appendix, we enlist the superconformal algebra and the associated generators in twistor space in terms of $\mathcal{Z}$ variables. The conformal algebra in three dimensions consists of the following generators: The generator of translations, $P_\mu$, the generator of rotations, $M_{\mu\nu}$, the generator of dilatation and the generator of special conformal transformations which are respectively denoted as $D$ and $K_\mu$. The supersymmetric extension to it admits the supercharge $Q_{a}^N$ and the generator of special superconformal transformations $S_{a}^N$ on top of the usual conformal algebra.
\begin{align}\label{SCFTalgebra}
[M_{\mu\nu},M_{\rho\lambda}]&=i\left(\delta_{\mu\rho}M_{\nu\lambda}-\delta_{\nu\rho}M_{\mu\lambda}-\delta_{\mu\lambda}M_{\nu\rho}+\delta_{\nu\lambda}M_{\mu\rho}\right),\notag\\
    [M_{\mu\nu},P_\alpha]&=i\left(\delta_{\mu\alpha}P_\nu-\delta_{\nu\alpha}P_\mu\right),\qquad\quad\qquad~~[M_{\mu\nu},K_\alpha]=i\left(\delta_{\mu\alpha}K_\nu-\delta_{\nu\alpha}K_\mu\right),\notag\\
    [D,P_\mu]&=i P_\mu,\qquad \qquad\qquad\qquad~\qquad\quad~~~~~[P_\mu,K_\nu]=2i\left(\delta_{\mu\nu}D-M_{\mu\nu}\right),\notag\\
    \{Q_{a}^A,Q_{b}^B\}&=(\sigma_\mu)_{ab}P^\mu\delta^{AB},\qquad\qquad~~~~~~~~~~~~~\{S_{a}^A,S_{b}^B\}=(\sigma_\mu)_{ab}K^\mu\delta^{AB},\notag\\
    [D,Q_{a}^A]&=\frac{i}{2}Q_{a}^A,\qquad\qquad\qquad\qquad~~~~~~~~~~~~~~[D,K_\mu]=-i K_\mu,\notag\\
    [K_\mu,Q_{a}^A]&=i(\sigma_\mu)_a^b S_{b}^A,\qquad\qquad\qquad~~~~~~~~~~~~~~~[D,S_{a}^A]=-\frac{i}{2}S_{a}^A,\notag\\
    [M_{\mu\nu},Q_{a}^A]&=\frac{i}{2}\epsilon_{\mu\nu\rho}(\sigma^\rho)^b_a Q_{b}^A,\qquad\quad\quad~~~~~~~~~~~~~~[P_\mu,S_{a}^A]=i(\sigma_\mu)^b_a Q_{b}^A,
    \notag\\
     [M_{\mu\nu},S_{a}^A]&=\frac{i}{2}\epsilon_{\mu\nu\rho}(\sigma^\rho)^b_a S_{b}^A,\quad\qquad\quad~~~~~~~~~~~~\{Q_{a}^A,S_{b}^B\}=\epsilon_{ab}D\delta^{AB}-\frac{i}{2}\epsilon_{\mu\nu\rho}(\sigma^\rho)_{ab}M^{\mu\nu}\delta^{AB}.
\end{align}
The (anti-)commutators not listed above are zero. The representation of these generators in $\mathcal{Z}=(\lambda,\bar{\mu},\psi)$, acting on primary supefields of $\Delta=2$ are as follows,
\begin{align}\label{SCFTGenZ}
\notag P_{ab} = i\;\lambda_{(a}\frac{\partial}{\partial\bar{\mu}^{b)}},\quad&\quad\notag K_{ab} = i\;\bar{\mu}_{(a}\frac{\partial}{\partial\lambda^{b)}},\\
\notag M_{ab}=i\Big(\bar{\mu}_{(a}\frac{\partial}{\partial\bar{\mu}^{b)}} + \lambda_{(a}\frac{\partial}{\partial\lambda^{b)}}\Big),\quad&\quad\notag D =\;\frac{i}{2} \left(\lambda^a\frac{\partial}{\partial\lambda^a}-\bar{\mu}^a \frac{\partial}{\partial\bar{\mu}^a} \right),\\  
Q_{a}^N=\;\frac{e^{-\frac{i\pi}{4}}}{\sqrt{2}}\bigg(\lambda_a\frac{\partial}{\partial\psi_{N}}+\psi^{N}\frac{\partial}{\partial\bar{\mu}^a}\bigg),\quad&\quad
S_{a}^N= \sqrt{2}e^{\frac{i\pi}{4}}\bigg(\bar{\mu}_a\frac{\partial}{\partial\psi_{N}}+\psi^{N}\frac{\partial}{\partial\lambda^a}\bigg).
\end{align}
The representation of these generators in $\mathcal{W}=(\bar{\lambda},\mu,\bar{\psi})$, acting on primary supefields of $\Delta=2$ are as follows,
\begin{align}\label{SCFTGenW}
\notag P_{ab} = -i\;\bar{\lambda}_{(a}\frac{\partial}{\partial\mu^{b)}},\quad&\quad\notag K_{ab} = -i\;\mu_{(a}\frac{\partial}{\partial\bar{\lambda}^{b)}},\\
\notag M_{ab}=i\Big(\mu_{(a}\frac{\partial}{\partial\mu^{b)}} + \bar{\lambda}_{(a}\frac{\partial}{\partial\bar{\lambda}^{b)}}\Big),\quad&\quad\notag D =\;\frac{i}{2} \left(\bar{\lambda}^a\frac{\partial}{\partial\bar{\lambda}^a}-\mu^a \frac{\partial}{\partial\mu^a} \right),\\  
Q_{a}^N=\;\frac{e^{-\frac{i\pi}{4}}}{\sqrt{2}}\bigg(\bar{\lambda}_a\frac{\partial}{\partial\bar{\psi}_{N}}-\bar{\psi}^{N}\frac{\partial}{\partial\mu^a}\bigg),\quad&\quad
S_{a}^N= \sqrt{2}e^{-\frac{i\pi}{4}}\bigg(\mu_a\frac{\partial}{\partial\bar{\psi}_{N}}-\bar{\psi}^{N}\frac{\partial}{\partial\bar{\lambda}^a}\bigg).
\end{align}

\section{Details on new Grassmann twistor variables}\label{app:gtv}
In this appendix, we present the necessary transformations required to go from momentum superspace variables to the Grassmann twistor variables. Recall \eqref{GTT} that we can perform a \textit{half-Fourier} transform with respect to either $\eta$ or $\bar{\eta}$ for $\mathbf{J}_s^{\pm}$:
\begin{align}\label{etatoxi2}
\mathbf{\Tilde{J}}_s^{\pm}(\bar{\chi},\bar{\eta})=\int d\eta~e^{-\frac{\bar{\chi}\eta}{4}}\mathbf{J}_s^{\pm}(\eta,\bar{\eta}),\quad\quad
\mathbf{\Tilde{J}}_s^{\pm}(\eta,\chi)=\int d\bar{\eta}~e^{-\frac{\chi\bar{\eta}}{4}}\mathbf{J}_s^{\pm}(\eta,\bar{\eta}).
\end{align}
Performing the integrals we obtain,
\begin{align}
\mathbf{\Tilde{J}}_s^{-}(\bar{\xi}_+,\bar{\xi}_-)=\frac{\bar{\xi}_-}{4}\bigg(J_s^{-}+\frac{\bar{\xi}_+}{4\sqrt{p}}J_{s+\frac{1}{2}}^{-}\bigg),\quad&\quad\mathbf{\Tilde{J}}_s^{+}(\bar{\xi}_+,\bar{\xi}_-)=\frac{1}{4}\bigg(\bar{\xi}_+ J_s^{+}+\frac{2}{\sqrt{p}}J_{s+\frac{1}{2}}^{+}\bigg),\notag\\
\mathbf{\Tilde{J}}_s^{-}(\xi_+,\xi_-)=\frac{1}{4}\bigg(\xi_+ J_s^{-}+\frac{2}{\sqrt{p}}J_{s+\frac{1}{2}}^{-}\bigg),\quad&\quad\mathbf{\Tilde{J}}_s^{+}(\xi_+,\xi_-)=\frac{\xi_-}{4}\bigg(J_s^{+}+\frac{\xi_+}{4\sqrt{p}}J_{s+\frac{1}{2}}^{+}\bigg).
\end{align}
where $\xi_\pm=\chi\pm\eta\And\bar{\xi}_\pm=\bar{\chi}\pm\bar{\eta}$.
Since the $\xi_{-},\bar{\xi}_-$ dependence is overall\footnote{We take a note of the fact that $\xi_-,\bar{\xi}_-$ do not show up anywhere in SCFT generators.}, let us integrate over them and define,
\begin{align}
    \mathbf{\hat{J}}_s^{-}=2\sqrt{2}\int d\bar{\xi}_{-}~\mathbf{\Tilde{J}}_s^{-}(\bar{\xi}_+,\bar{\xi}_-),\quad\quad\mathbf{\hat{J}}_s^{+}=2\sqrt{2}\int d{\xi_{-}}~\mathbf{\Tilde{J}}_s^{+}(\xi_+,\xi_-).
\end{align}
Since $\xi_-,\bar{\xi}_-$ has completely dropped out of the definition of the currents, we now perform one final redefinition of variable with respect to $\xi_+,\bar{\xi}_+$,
\begin{align}\label{newgrassman}
\psi=\frac{e^{-\frac{i\pi}{4}}}{2\sqrt{2}}\xi_+,\quad\quad\bar{\psi}=\frac{e^{-\frac{i\pi}{4}}}{2\sqrt{2}}\bar{\xi}_+.
\end{align}
Finally, our supercurrents after rescaling with a factor of $\sqrt{2}$ take the following form,
\begin{align}
\mathbf{\hat{J}}_s^{-}(\bar{\psi})=J_s^{-}+\frac{e^{\frac{i\pi}{4}}\bar{\psi}}{\sqrt{2p}}J_{s+\frac{1}{2}}^{-},\quad&\quad
\mathbf{\hat{J}}_s^{+}(\bar{\psi})=e^{\frac{i\pi}{4}}\bar{\psi}J_s^{+}+\frac{1}{\sqrt{2p}}J_{s+\frac{1}{2}}^{+},\notag\\
\mathbf{\hat{J}}_s^{-}(\psi)=e^{\frac{i\pi}{4}}\psi J_s^{-}+\frac{1}{\sqrt{2p}}J_{s+\frac{1}{2}}^{-},\quad&\quad
\mathbf{\hat{J}}_s^{+}(\psi)=J_s^{+}+\frac{e^{\frac{i\pi}{4}}\psi}{\sqrt{2p}}J_{s+\frac{1}{2}}^{+}.
\end{align}
The action of the supersymmetry generator in spinor-helicity variables and the new Grassmann twistor variables is found to be\footnote{This can be derived by tracing through the steps required to go from the $(\eta,\bar{\eta})$ variables to these $(\psi,\bar{\psi})$ ones.},
\begin{align}\label{QSH}
[Q_a,\mathbf{\hat{J}}_s^{\pm}(\bar{\psi})]=-\frac{e^{-\frac{i\pi}{4}}}{\sqrt{2}}\bigg(\bar{\lambda}_{a}\frac{\partial}{\partial\bar{\psi}}+i\lambda_a\bar{\psi}\bigg)\mathbf{\hat{J}}_s^{\pm}(\bar{\psi}),\quad\quad[Q_a,\mathbf{\hat{J}}_s^{\pm}(\psi)]=\frac{e^{-\frac{i\pi}{4}}}{\sqrt{2}}\bigg(\lambda_{a}\frac{\partial}{\partial\psi}+i\bar{\lambda}_a\psi\bigg)\mathbf{\hat{J}}_s^{\pm}(\psi).
\end{align}
Using \eqref{QSH}, the supersymmetric Ward-identity for $n-$point functions is as follows,
\begin{align}
    \sum_{i=1}^{n} \big\langle0| \mathbf{\hat{J}}_{s_1}(\psi_{1})\cdots \big[Q_{a},\mathbf{\hat{J}}_{s_i}(\psi_i)\big] 
    \cdots \mathbf{\hat{J}}_{s_n}(\bar{\psi}_n)\big|0\rangle=0.
\end{align}

\section{Integral representation of delta function}\label{app:delta}
In this appendix, we give the details of integral representation of delta function which comes handy for packaging the component correlators in a manifest supertwistor manner.\\
The integral representation of delta-function can be viewed as an integral over Schwinger parameter,
\begin{align}
\delta^{[n]}(x)\equiv i^{n}\int\frac{dc}{2\pi}(c^nexp{(icx)})\qquad\qquad n\in Z_{\geq0},
\end{align}
where $\delta^{[n]}(x)$ denotes the $n^{th}$ derivative of the delta-function.
Recall the three point function for the +++ helicity configuration in \eqref{eqn4.15}. We present it again for the convenience of the reader.
\begin{align}
&\notag\langle0| \mathbf{\hat{J}_{s_{1}}^+}(\mathcal{Z}_1) \mathbf{\hat{J}_{s_{2}}^+}(\mathcal{Z}_2) \mathbf{\hat{J}_{s_{3}}^+}(\mathcal{Z}_3)|0\rangle \\ \notag
&= (-i)^{(s_1+s_2+s_3)}c_{s_1s_2s_3}\Big(\delta^{[s_1+s_2-s_3]}(Z_{1}\cdot Z_{2})\delta^{[s_2+s_3-s_1]}(Z_{2}\cdot Z_{3})\delta^{[s_3+s_1-s_2]}(Z_{3}\cdot Z_{1})  \\ \notag
&+\;\psi_{1}\psi_{2}\; \delta^{[s_1+s_2+1-s_3]}(Z_{1}\cdot Z_{2})\delta^{[s_2+s_3-s_1]}(Z_{2}\cdot Z_{3})\delta^{[s_3+s_1-s_2]}(Z_{3}\cdot Z_{1}) \\ \notag 
&+\;\psi_{2}\psi_{3}\; \delta^{[s_1+s_2-s_3]}(Z_{1}\cdot Z_{2})\delta^{[s_2+s_3+1-s_1]}(Z_{2}\cdot Z_{3})\delta^{[s_3+s_1-s_2]}(Z_{3}\cdot Z_{1}) \\ \notag
&+\;\psi_{3}\psi_{1}\; \delta^{[s_1+s_2-s_3]}(Z_{1}\cdot Z_{2})\delta^{[s_2+s_3-s_1]}(Z_{2}\cdot Z_{3})\delta^{[s_3+s_1+1-s_2]}(Z_{3}\cdot Z_{1})\Big).  
\end{align}
Now using the integral representation of delta-function, it can be rewritten as,
\begin{align}
\notag\langle 0| &\mathbf{\hat{J}_{s_{1}}^+}(\mathcal{Z}_1)  \mathbf{\hat{J}_{s_{2}}^+}(\mathcal{Z}_2) \mathbf{\hat{J}_{s_{3}}^+}(\mathcal{Z}_3)|0\rangle \\ \notag  
&=c_{s_1s_2s_3} \int \frac{d^3c_{ij}}{(2\pi)^3}\;e^{(ic_{12}Z_1\cdot Z_2+cyc)}(1-\textit{i} c_{12}\psi_1\psi_2-\textit{i} c_{23}\psi_2\psi_3-\textit{i} c_{31}\psi_3\psi_1) c_{12}^{s_1+s_2-s_3}c_{23}^{s_2+s_3-s_1}c_{31}^{s_3+s_1-s_2}\\ \notag  &=c_{s_1s_2s_3}\int \frac{d^3c_{ij}}{(2\pi)^3}\;e^{(ic_{12}Z_1\cdot Z_2+cyc)}e^{(-ic_{12}\psi_1\psi_2-ic_{23}\psi_2\psi_3-ic_{31}\psi_3\psi_1)}c_{12}^{s_1+s_2-s_3}c_{23}^{s_2+s_3-s_1}c_{31}^{s_3+s_1-s_2}\\
\notag&=c_{s_1s_2s_3}\int \frac{d^3c_{ij}}{(2\pi)^3}\;e^{(ic_{12}(Z_1\cdot Z_2-\psi_1\psi_2)+cyc)}c_{12}^{s_1+s_2-s_3}c_{23}^{s_2+s_3-s_1}c_{31}^{s_3+s_1-s_2}\\
\notag&=c_{s_1s_2s_3}\int \frac{d^3c_{ij}}{(2\pi)^3}\;e^{(ic_{12}\mathcal{Z}_1\cdot \mathcal{Z}_2+cyc)}c_{12}^{s_1+s_2-s_3}c_{23}^{s_2+s_3-s_1}c_{31}^{s_3+s_1-s_2}\\
&=c_{s_1s_2s_3}(-\textit{i})^{s_1+s_2+s_3}\delta^{[s_1+s_2-s_3]}(\mathcal{Z}_{1}\cdot \mathcal{Z}_{2})\delta^{[s_2+s_3-s_1]}(\mathcal{Z}_{2}\cdot \mathcal{Z}_{3})\delta^{[s_3+s_1-s_2]}(\mathcal{Z}_{3}\cdot \mathcal{Z}_{1}),
\end{align}
where, the supertwistor dot products are the same as defined in \eqref{supertwistorsdotExt}.
\section{Super Fourier transform of helicity operator}\label{app:Fourier}
In this section we calculate the super-Fourier transform of the helicity operator \eqref{superhelicity1} from the $\mathcal{Z}$ variable to the dual supertwistor variable $\mathcal{W}$. The steps are as follows,

\begin{align*}
    \mathbf{h}(\mathcal{Z})f(\mathcal{Z}) &= -\frac{1}{2}\left( \mathcal{Z}_{\mathcal{A}}\frac{\partial}{\partial\mathcal{Z}_{\mathcal{A}}}+2 \right)f(\mathcal{Z})\\
    &= -\frac{1}{2}\left( \mathcal{Z}_{\mathcal{A}}\frac{\partial}{\partial\mathcal{Z}_{\mathcal{A}}}+2 \right)\int d^{4|\mathcal{N}}\mathcal{W} e^{i\mathcal{Z}\cdot\mathcal{W}}\tilde{f}(\mathcal{W})\\
    &= -\frac{1}{2}\int d^{4|\mathcal{N}}\mathcal{W}\left( \mathcal{Z}_{\mathcal{A}}\frac{\partial}{\partial\mathcal{Z}_{\mathcal{A}}}e^{i\mathcal{Z}\cdot\mathcal{W}}\tilde{f}(\mathcal{W}) +2e^{i\mathcal{Z}\cdot\mathcal{W}}\tilde{f}(\mathcal{W}) \right)\\
    & = -\frac{1}{2}\int d^{4|\mathcal{N}}\mathcal{W}\left( i\mathcal{Z}_{\mathcal{A}}\mathcal{W}^{\mathcal{A}}e^{i\mathcal{Z}\cdot\mathcal{W}}\tilde{f}(\mathcal{W}) +2e^{i\mathcal{Z}\cdot\mathcal{W}}\tilde{f}(\mathcal{W}) \right)\\
    & = -\frac{1}{2}\int d^{4|\mathcal{N}}\mathcal{W}\left( (-1)^{\eta_{\mathcal{A}}\eta_{\mathcal{B}}} \Omega^{\mathcal{AB}} i\mathcal{W}_{\mathcal{B}}\mathcal{Z}_{\mathcal{A}}e^{i\mathcal{Z}\cdot\mathcal{W}}\tilde{f}(\mathcal{W}) +2e^{i\mathcal{Z}\cdot\mathcal{W}}\tilde{f}(\mathcal{W}) \right)\\
    & = -\frac{1}{2}\int d^{4|\mathcal{N}}\mathcal{W}\left( \mathcal{W}_{\mathcal{B}}\frac{\partial}{\partial \mathcal{W}_{\mathcal{B}}}e^{i\mathcal{Z}\cdot\mathcal{W}}\tilde{f}(\mathcal{W}) +2e^{i\mathcal{Z}\cdot\mathcal{W}}\tilde{f}(\mathcal{W}) \right)\\
    & = -\frac{1}{2}\int d^{4|\mathcal{N}}\mathcal{W}\left( -(-1)^{\eta_{\mathcal{B}}\eta_{\mathcal{C}}}\delta^{\mathcal{B}}_{\mathcal{C}}\delta^{\mathcal{C}}_{\mathcal{B}} - \mathcal{W}_{\mathcal{B}}\frac{\partial}{\partial \mathcal{W}_{\mathcal{B}}} +2 \right)\tilde{f}(\mathcal{W})e^{i\mathcal{Z}\cdot\mathcal{W}}\\
    & = \frac{1}{2}\int d^{4|\mathcal{N}}\mathcal{W}\left( (-1)^{\eta_{\mathcal{B}}\eta_{\mathcal{C}}}\delta^{\mathcal{B}}_{\mathcal{C}}\delta^{\mathcal{C}}_{\mathcal{B}} + \mathcal{W}_{\mathcal{B}}\frac{\partial}{\partial \mathcal{W}_{\mathcal{B}}} - 2 \right)\tilde{f}(\mathcal{W})e^{i\mathcal{Z}\cdot\mathcal{W}}\\
    & = \frac{1}{2}\int d^{4|\mathcal{N}}\mathcal{W}\left( (4-\mathcal{N}) + \mathcal{W}_{\mathcal{B}}\frac{\partial}{\partial \mathcal{W}_{\mathcal{B}}} - 2 \right)\tilde{f}(\mathcal{W})e^{i\mathcal{Z}\cdot\mathcal{W}}\\
    & = \frac{1}{2}\int d^{4|\mathcal{N}}\mathcal{W}\left( (2-\mathcal{N}) + \mathcal{W}_{\mathcal{B}}\frac{\partial}{\partial \mathcal{W}_{\mathcal{B}}} \right)\tilde{f}(\mathcal{W})e^{i\mathcal{Z}\cdot\mathcal{W}}.
\end{align*}
We can read off the helicity operator in $\mathcal{W}$ variables to be,
\begin{align}\label{ManifestDualHelicity}
    \mathbf{h}(\mathcal{W}) = \frac{1}{2}\left( \mathcal{W}_{\mathcal{B}}\frac{\partial}{\partial \mathcal{W}_{\mathcal{B}}} + 2 \right) - \frac{\mathcal{N}}{2}.
\end{align}

\section{Details on solving superconformal Ward identity}\label{app:WI}
In this appendix, we present the details of solving the superconformal Ward identity in manifest supertwistor space for two point function. Recall the Ward identity equation \eqref{ManifestSCFTWI},
\begin{align}\label{T_ABsupertwistorgenerator}
&\left( \mathcal{Z}_1^{(\mathcal{A}}\frac{\partial}{\partial \mathcal{Z}_{1\mathcal{B}]}} + \mathcal{Z}_2^{(\mathcal{A}}\frac{\partial}{\partial \mathcal{Z}_{2\mathcal{B}]}} \right)\langle 0|\hat{\mathbf{J}}_s^+(\mathcal{Z}_1)\hat{\mathbf{J}}_s^+(\mathcal{Z}_2)|0\rangle = 0\notag\\
\implies &\sum_{i=1}^{2}\left( \mathcal{Z}_i^\mathcal{A}\frac{\partial}{\partial\mathcal{Z}_{i\mathcal{B}}} + (-1)^{\eta_\mathcal{A}\eta_\mathcal{B}}\mathcal{Z}_i^\mathcal{B}\frac{\partial}{\partial\mathcal{Z}_{i\mathcal{A}}} \right)\langle 0|\hat{\mathbf{J}}_s^+(\mathcal{Z}_1)\hat{\mathbf{J}}_s^+(\mathcal{Z}_2)|0\rangle = 0.
\end{align}
In order to solve the above equation, let us take the projection of the above equation on three independent basis $\mathcal{Z}_{1\mathcal{B}}\mathcal{Z}_{1\mathcal{A}}, \mathcal{Z}_{2\mathcal{B}}\mathcal{Z}_{2\mathcal{A}}$ and $\mathcal{Z}_{1\mathcal{B}}\mathcal{Z}_{2\mathcal{A}}$.
So, taking dot product of \eqref{T_ABsupertwistorgenerator} with $\mathcal{Z}_{1\mathcal{B}}\mathcal{Z}_{1\mathcal{A}}$ and setting $\mathcal{Z}_i\cdot\mathcal{Z}_i = 0$, we get,
\begin{align}\label{step1}
&\mathcal{Z}_{1\mathcal{B}}\mathcal{Z}_{1\mathcal{A}}\sum_{i=1}^{2} \left( \mathcal{Z}_i^\mathcal{A}\frac{\partial}{\partial\mathcal{Z}_{i\mathcal{B}}}+(-1)^{\eta_\mathcal{A}\eta_\mathcal{B}}\mathcal{Z}_i^\mathcal{B}\frac{\partial}{\partial\mathcal{Z}_{i\mathcal{A}}}  \right)\langle 0|\hat{\mathbf{J}}_s^+(\mathcal{Z}_1)\hat{\mathbf{J}}_s^+(\mathcal{Z}_2)|0\rangle = 0 \notag\\
\implies & \left( \mathcal{Z}_{1\mathcal{A}} \mathcal{Z}_2^\mathcal{A} \mathcal{Z}_{1\mathcal{B}} \frac{\partial}{\partial\mathcal{Z}_{2\mathcal{B}}} + (-1)^{2\eta_\mathcal{A}\eta_\mathcal{B}}\mathcal{Z}_{1\mathcal{B}}\mathcal{Z}_2^\mathcal{B}\mathcal{Z}_{1\mathcal{A}}\frac{\partial}{\partial\mathcal{Z}_{2\mathcal{A}}} \right)\langle 0|\hat{\mathbf{J}}_s^+(\mathcal{Z}_1)\hat{\mathbf{J}}_s^+(\mathcal{Z}_2)|0\rangle = 0 \notag\\
\implies &2(\mathcal{Z}_1\cdot\mathcal{Z}_2)\left( \mathcal{Z}_1\cdot\frac{\partial}{\partial\mathcal{Z}_2} \right)\langle 0|\hat{\mathbf{J}}_s^+(\mathcal{Z}_1)\hat{\mathbf{J}}_s^+(\mathcal{Z}_2)|0\rangle = 0\notag\\
\implies&\left( \mathcal{Z}_1\cdot\frac{\partial}{\partial\mathcal{Z}_2} \right)\langle 0|\hat{\mathbf{J}}_s^+(\mathcal{Z}_1)\hat{\mathbf{J}}_s^+(\mathcal{Z}_2)|0\rangle = 0\qquad\qquad(\textrm{Since} \;(\mathcal{Z}_1\cdot\mathcal{Z}_2) \neq 0)\notag\\
\implies &\frac{\partial}{\partial \mathcal{Z}_{2\mathcal{A}}}\langle 0|\hat{\mathbf{J}}_s^+(\mathcal{Z}_1)\hat{\mathbf{J}}_s^+(\mathcal{Z}_2)|0\rangle = \alpha \mathcal{Z}_1^\mathcal{A}.
\end{align}
Similarly, taking dot product of \eqref{T_ABsupertwistorgenerator} with $\mathcal{Z}_{2\mathcal{B}}\mathcal{Z}_{2\mathcal{A}}$, we get,
\begin{align}\label{step2}
&\left( \mathcal{Z}_2\cdot\frac{\partial}{\partial\mathcal{Z}_1} \right)\langle 0|\hat{\mathbf{J}}_s^+(\mathcal{Z}_1)\hat{\mathbf{J}}_s^+(\mathcal{Z}_2)|0\rangle = 0\notag\\
\implies& \frac{\partial}{\partial \mathcal{Z}_{1\mathcal{A}}}\langle 0|\hat{\mathbf{J}}_s^+(\mathcal{Z}_1)\hat{\mathbf{J}}_s^+(\mathcal{Z}_2)|0\rangle = \beta \mathcal{Z}_2^\mathcal{A}.
\end{align}
Finally, taking dot product of \eqref{T_ABsupertwistorgenerator} with $\mathcal{Z}_{1\mathcal{B}}\mathcal{Z}_{2\mathcal{A}}$, we get,
\begin{align}\label{step3}
(\mathcal{Z}_2\cdot\mathcal{Z}_1)\mathcal{Z}_1\cdot\frac{\partial \langle 0|\hat{\mathbf{J}}_s^+(\mathcal{Z}_1)\hat{\mathbf{J}}_s^+(\mathcal{Z}_2)|0\rangle}{\partial \mathcal{Z}_1} + (\mathcal{Z}_1\cdot\mathcal{Z}_2)\mathcal{Z}_2\cdot\frac{\partial \langle 0|\hat{\mathbf{J}}_s^+(\mathcal{Z}_1)\hat{\mathbf{J}}_s^+(\mathcal{Z}_2)|0\rangle}{\partial \mathcal{Z}_2} = 0.
\end{align}
Substituting \eqref{step1} and \eqref{step2} in \eqref{step3}, we have $\alpha = -\beta$ and \eqref{step1}, \eqref{step2} implies that the two point function being scalar has the argument,
\begin{align}
\langle 0|\hat{\mathbf{J}}_s^+(\mathcal{Z}_1)\hat{\mathbf{J}}_s^+(\mathcal{Z}_2)|0\rangle = F(\mathcal{Z}_1\cdot\mathcal{Z}_2).
\end{align}

\section{Details of solving helicity equation}\label{app:helicity}
In this appendix, we present the details of solving the helicity differential equation in manifest supertwistor space for (++) super-correlator. Recall the helicity identity \eqref{superhelicityaction2point} for two point functions,
\begin{align}
 &\mathcal{Z}_{1 \mathcal{A}} \frac{\partial}{\partial \mathcal{Z}_{1 \mathcal{A}}} F \left( \mathcal{Z}_1 \cdot
\mathcal{Z}_2 \right) = -2 \left(s + 1\right) F \left( \mathcal{Z}_1 \cdot
\mathcal{Z}_2 \right)\notag\\
\implies &\mathcal{Z}_{1 \mathcal{A}} \frac{\partial \left( \mathcal{Z}_1 \cdot
\mathcal{Z}_2\right)}{\partial \mathcal{Z}_{1 \mathcal{A}}} \frac{\partial F \left(\mathcal{Z}_1 \cdot
\mathcal{Z}_2\right)}{\partial \left(\mathcal{Z}_1 \cdot
\mathcal{Z}_2\right)}  = -2 \left(s + 1\right) F \left( \mathcal{Z}_1 \cdot
\mathcal{Z}_2 \right) \notag\\
\implies &  \mathcal{Z}_{1 \mathcal{A}} \mathcal{Z}_2^\mathcal{A} \frac{\partial F \left(\mathcal{Z}_1 \cdot
\mathcal{Z}_2\right)}{\partial \left(\mathcal{Z}_1 \cdot
\mathcal{Z}_2\right)} =   -2 \left(s + 1\right) F \left( \mathcal{Z}_1 \cdot
\mathcal{Z}_2 \right)\notag\\
\implies & \mathcal{Z}_1 \cdot \mathcal{Z}_2 \frac{d F \left(\mathcal{Z}_1 \cdot
\mathcal{Z}_2\right)}{d \left(\mathcal{Z}_1 \cdot
\mathcal{Z}_2\right)} =   -2 \left(s + 1\right) F \left( \mathcal{Z}_1 \cdot
\mathcal{Z}_2 \right)\notag\\
\implies & \int \frac{d F}{F} = -2 \left(s + 1\right) \int \frac{d \left( \mathcal{Z}_1 \cdot \mathcal{Z}_2 \right)}{\mathcal{Z}_1 \cdot \mathcal{Z}_2 }\notag\\
\implies & \text{ln} (F) = -2 \left(s + 1\right) \text{ln} \left( \mathcal{Z}_1 \cdot \mathcal{Z}_2\right)+\text{ln}(c_s)\notag\\
\implies &F \left( \mathcal{Z}_1 \cdot \mathcal{Z}_2\right) = \frac{c_s}{\left( \mathcal{Z}_1 \cdot \mathcal{Z}_2\right)^{2\left(s+1\right)}}.
\end{align}
The solving of above differential equation yields the classical solution. However, there is another class of solutions, namely weak solutions (distributional), for which the analysis is discussed in detail in \cite{Bala:2025gmz}. One can perform a similar computation for the three point functions.

\section{Reality conditions and CPT invariance in supertwistor space}\label{app:realitysusy}
In this appendix, our aim is twofold: First, we shall discuss certain important reality conditions in supertwistor space: Then, we shall briefly discuss the implication of $CPT$ invariance on the super-correlators.
\subsection{Reality conditions}
Let us first start with the reality properties of the different variables we have used in this paper. To begin with, consider the spinor helicity variables $(\lambda,\Bar{\lambda})$. They are real for spacelike momenta \cite{Baumann:2024ttn,Bala:2025gmz} i.e. $\lambda^*=\lambda$ and $\bar{\lambda}^*=\bar{\lambda}$. Therefore, the twistor and dual twistor variables $Z^A=(\lambda^a,\Bar{\mu}_{a'})$ and $W_A=(\mu_a,\Bar{\lambda}^a)$ obtained via half-Fourier transforms are real as well. Moving on to the Grassmann coordinates, the reality of the spinor helicity variables implies that so are the Grassmann variables $\eta,\Bar{\eta}$ through the definition \eqref{SHandTheta}. We then followed by a Grassmann twistor transform and landed up with real Grassmann variables $\xi_{\pm},\Bar{\xi}_{\pm}$ in \eqref{etatoxi2} which are coordinates on the Grassmann twistor space. Till this step, there are no complex variables. However, we introduced the variables $\psi$ and $\Bar{\psi}$ in \eqref{newgrassman}. These variables are complex and satisfy the reality conditions,
\begin{align}\label{grassreal}
    \psi^*=i \psi,\Bar{\psi}^*=i\Bar{\psi}.
\end{align}
 Let us  now discuss the reality properties of super-correlators. First of all, in spinor helicity variables the (parity-even) correlators are purely real and thus,
\begin{align}
    \langle 0|\prod_{i=1}^{n}J_{s_i}(\lambda_i,\Bar{\lambda}_i,\psi)|0\rangle^*=\langle 0|\prod_{i=1}^{n}J_{s_i}(\lambda_i,\Bar{\lambda}_i,\psi)|0\rangle.
\end{align}
Using the half-Fourier transform, the associated conditions in twistor space becomes,
\begin{align}\label{realityofsusytwistor}
     \langle 0|\prod_{i=1}^{n}J_{s_i}(\lambda_i,\Bar{\mu}_i,\psi)|0\rangle^*=\langle 0|\prod_{i=1}^{n}J_{s_i}(\lambda_i,-\Bar{\mu}_i,\psi)|0\rangle.
\end{align}
The analogous statement in dual-twistor space is obvious.
Now, given \eqref{realityofsusytwistor}, let us check that our correlators satisfy this condition. To this end, we consider the different super twistor dot products \eqref{supertwistorsdot1} that make up the correlators. For example consider,
\begin{align}
    \mathcal{Z}_i\cdot\mathcal{Z}_j=Z_i\cdot Z_j-\psi_i \psi_j.
\end{align}
Taking a complex conjugate we get,
\begin{align}
    (\mathcal{Z}_i\cdot\mathcal{Z}_j)^*=(Z_i\cdot Z_j)^*-(\psi_i\cdot \psi_j)^*=Z_i\cdot Z_j-\psi_j^*\cdot \psi_i^*=Z_i\cdot Z_j+\psi_j\psi_i=Z_i\cdot Z_j-\psi_i\cdot \psi_j=\mathcal{Z}_i\cdot \mathcal{Z}_j.
\end{align}
 Similarly, we have,
\begin{align}
    (\mathcal{Z}_i\cdot\mathcal{W}_j)^*=(Z_i\cdot W_j)^*+(\psi_i\cdot \Bar{\psi}_j)^*=Z_i\cdot W_j+\Bar{\psi}_j^*\cdot \psi_i^*=Z_i\cdot Z_j-\Bar{\psi}_j\psi_i=Z_i\cdot W_j+\psi_i\cdot\Bar{\psi}_j=\mathcal{Z}_i\cdot\mathcal{W}_j.
\end{align}
Consider a concrete example of a parity even two point function in the $(++)$ helicity \eqref{12+}. We have,
\begin{align}\label{H.7}
    \langle 0|\mathbf{\hat{J}}_{s}^+(\lambda_1,\Bar{\mu}_1,\psi_1)\mathbf{\hat{J}}_{s}^+(\lambda_1,\Bar{\mu}_2,\psi_2)|0\rangle=\frac{i^{2s+2}}{(\mathcal{Z}_1\cdot \mathcal{Z}_2)^{2s+2}}=\frac{i^{2s+2}}{(\lambda_1\cdot \Bar{\mu}_2-\lambda_2\cdot \Bar{\mu}_1-\psi_1\psi_2)}.
\end{align}
Thus using \eqref{H.7} we have,
\begin{align}
    \langle 0|\mathbf{\hat{J}}_{s}^+(\lambda_1,\Bar{\mu}_1,\psi_1)\mathbf{\hat{J}}_{s}^+(\lambda_1,\Bar{\mu}_2,\psi_2)|0\rangle^*&=(-1)^{2s+2}\frac{i^{2s+2}}{(\lambda_1\cdot \Bar{\mu}_2-\lambda_2\cdot \Bar{\mu}_1-\psi_1\psi_2)} \notag \\ 
    &=\langle 0|\mathbf{\hat{J}}_{s_1}(\lambda_1,-\Bar{\mu}_1,\psi_1)\mathbf{\hat{J}}_{s_2}(\lambda_1,-\Bar{\mu}_2,\psi_2)|0\rangle,
\end{align}
which is the previously mentioned reality condition.
Therefore, all parity even correlation functions (consider a fixed helicity) are real and parity odd correlation functions are purely imaginary as can be seen for instance in \eqref{ManifestTwoPoint}.

Next, let us show that the full super Fourier transform preserves the reality properties of the currents. We show this via an illustrative example of the $\mathcal{N}=1$ case.
\begin{align}\label{ZtoWfullFT}
    \mathbf{\hat{J}}_s^{+}(\mathcal{W})=\int d^{4|1}\mathcal{Z}e^{-i\mathcal{Z}\cdot\mathcal{W}}\mathbf{\hat{J}}_s^{+}(\mathcal{Z}).
\end{align}
The important part here is the explicit form of the measure. Recalling \eqref{grassreal} implies the quantity $e^{\frac{i\pi}{4}}\psi$ is real. Consider,
\begin{align}
    \int d\psi \psi=1\implies \int d(e^{\frac{i\pi}{4}}\psi)e^{\frac{i\pi}{4}}\psi=1.
\end{align}
Thus the measure must scale as follows,
\begin{align}
    d(e^{\frac{i\pi}{4}}\psi)=e^{-\frac{i\pi}{4}}d\psi,
\end{align}
in order to preserve the value of the integral.
Coming back to the super Fourier transform we see that the correct real measure is,
\begin{align}
    d^{4|1}\mathcal{Z}=d^4 Z e^{-\frac{i\pi}{4}}d\psi.
\end{align}
This generalizes in a straightforward way to higher supersymmetry where the measure is given by,
\begin{align}
    d^{4|\mathcal{N}}\mathcal{Z}=d^4 Z e^{-\frac{i\mathcal{N}\pi}{4}}d^{\mathcal{N}}\psi.
\end{align}
Using this measure, it is easy to see that the reality of $\mathbf{\hat{J}_s}^{+}(\mathcal{Z})$ implies the reality of $\mathbf{\hat{J}_s}^{+}(\mathcal{W})$ in \eqref{ZtoWfullFT}.
\subsection{CPT invariance for the twistor correlator}
The implications of CPT invariance for each component correlator was discussed in \cite{Bala:2025gmz}. For example, the existence of the $(---)$ helicity three point function together with $CPT$ invariance implies that there must exist an appropriate $(+++)$ three point function.
In this appendix, we shall use this fact applied to supertwistor space correlators. One important fact in the analysis is that supersymmetry necessitates half integer spin currents and these require a more careful treatment. Before we proceed, we remind the reader the operations of parity, time-reversal and $PT$ on the spinor helicity variables\footnote{Note that $C$ is unimportant for us since all our correlators are even under $C$. See appendix E of \cite{Baumann:2024ttn} for more details.}. 
\begin{align}\label{PTSH}
    P&:(\lambda_1,\lambda_2)\to (-\Bar{\lambda}_1,\Bar{\lambda}_2)~,~(\Bar{\lambda}_1,\Bar{\lambda}_2)\to (-\lambda_1,\lambda_2),\notag\\
    T&:(\lambda_1,\lambda_2)\to (\lambda_2 -\lambda_1)~,~(\Bar{\lambda}_1,\Bar{\lambda}_2)\to (\Bar{\lambda}_2,-\Bar{\lambda}_1),\notag\\
    PT&:(\lambda_1,\lambda_2)\to-(\Bar{\lambda}_2,\Bar{\lambda}_1),(\Bar{\lambda}_1,\Bar{\lambda}_2)\to -(\lambda_2,\lambda_1).
\end{align}
In the superspace, we also need to specify the $PT$ properties of $\psi^A,\Bar{\psi}^{A}$. We find that their transformations are,
\begin{align}\label{PTGrassmann}
    &PT:\;\psi^A\to i \Bar{\psi}^{A}=\Bar{\psi}^{*},\;\Bar{\psi}^{A}\to i \psi^A=\psi^{*}.
\end{align}
We focus only on two point functions in detail here as the analysis easily generalizes to higher points.
\subsubsection{Two point functions:~I}
Let us begin with component level integer spin two point functions. The $(++)$ helicity two point function in twistor space is given by,
\begin{align}
    \langle 0|\mathbf{\hat{J}}_s^{+}(Z_1)\mathbf{\hat{J}}_s^{+}(Z_2)|0\rangle=(c_e+ic_o)\frac{i^{2s+2}}{(Z_1\cdot Z_2)^{2s+2}}.
\end{align}
By a half-Fourier transform to spinor helicity variables we get up to an overall factor \cite{Bala:2025gmz},
\begin{align}
    \int d^2\Bar{\mu}_1d^2\Bar{\mu}_2~e^{-i\Bar{\lambda}_1\cdot \Bar{\mu}_1-i\Bar{\lambda}_2\cdot \Bar{\mu}_2}\frac{(c_e+ic_o)i^{2s+2}}{(Z_1\cdot Z_2)^{2s+2}}=(c_e+ic_o)\frac{\langle \Bar{1}\Bar{2}\rangle^{2s}}{p_1^{2s-1}}\delta^{3}(p_1+p_2).
\end{align}
Performing $CPT$ on the entire equation one obtains the equation,
\begin{align}
    \int d^2 \mu_1 d^2\mu_2~e^{i\lambda_1\cdot \mu_1+i\lambda_2\cdot \mu_2}\frac{(c_e-ic_o)(-1)^{2s}i^{2s+2}}{(W_1\cdot W_2)^{2s+2}}=(-1)^{2s}(c_e-ic_o)\frac{\langle 1 2\rangle^{2s}}{p_1^{2s-1}}\delta^3(p_1+p_2).
\end{align}
This allows us to identify the $(--)$ twistor space two point function as,
\begin{align}
    \langle 0|\mathbf{\hat{J}}_s^{-}(W_1)\mathbf{\hat{J}}_s^{-}(W_2)|0\rangle=(-1)^{2s}(c_e-ic_o)\frac{i^{2s+2}}{(W_1\cdot W_2)^{2s+2}}.
\end{align}
In \cite{Bala:2025gmz}, this factor of $(-1)^{2s}$ was ignored as it is equal to $1$ for integer spins. It however, provides the $-1$ factor for half integer spins. For example, consider the $s=\frac{1}{2}$ case and focus on the parity even sector. In the two helicities we have,
\begin{align}
    &\langle 0|O_{1/2}^{-}(p_1)O_{1/2}^{-}(p_2)|0\rangle=c_e \frac{\langle 1 2\rangle}{p_1},\notag\\
    &\langle 0|O_{1/2}^{+}(p_1)O_{1/2}^{+}(p_2)|0\rangle=-c_e \frac{\langle \Bar{1} \Bar{2}\rangle}{p_1},
\end{align}
which under $CPT$ go to each other correctly.

\subsubsection{Two point functions:~II}
We now repeat the above analysis for the case when $-$ helicity currents are expressed in twistor space $+$ helicity currents are in dual twistor space. Our starting point is,
\begin{align}
    \langle 0|\mathbf{\hat{J}}_s^{-}(Z_1)\mathbf{\hat{J}}_s^{-}(Z_2)|0\rangle=(-1)^{2s}(c_e-ic_o)\frac{i^{-2s+2}}{(Z_1\cdot Z_2)^{-2s+2}}.
\end{align}
Performing a half-Fourier transform we get,
\begin{align}
    \int d^2\Bar{\mu}_1d^2\Bar{\mu}_2~e^{-i\Bar{\lambda}_1\cdot\Bar{\mu}_1-i\Bar{\lambda}_2\cdot \Bar{\mu}_2}(-1)^{2s}(c_e-ic_o)\frac{i^{-2s+2}}{(Z_1\cdot Z_2)^{-2s+2}}=(-1)^{2s}(c_e-ic_o)\frac{\langle 1 2\rangle^{2s}}{p_1^{2s-1}}\delta^3(p_1+p_2).
\end{align}
Performing a CPT transformation we get,
\begin{align}
    \int d^2\mu_1d^2\mu_2~e^{i\lambda_1\cdot\mu_1+i\lambda_2\cdot \mu_2}(c_e+ic_o)\frac{i^{-2s+2}}{(W_1\cdot W_2)^{-2s+2}}=(c_e+ic_o)\frac{\langle \Bar{1} \Bar{2}\rangle^{2s}}{p_1^{2s-1}}\delta^3(p_1+p_2).
\end{align}
This tells us that the $(++)$ helicity two point function is given by,
\begin{align}
    \langle 0|\mathbf{\hat{J}}_s^{+}(W_1)\mathbf{\hat{J}}_s^{+}(W_2)|0\rangle=(c_e+ic_o)\frac{i^{-2s+2}}{(W_1\cdot W_2)^{-2s+2}}.
\end{align}
\subsubsection{Summary for two points}
Thus, we see that under $PT$,
\begin{align}
    &PT:(c_e+ic_o)\frac{i^{2s+2}}{(Z_1\cdot Z_2)^{2s+2}}\to (-1)^{2s}(c_e-ic_o)\frac{i^{2s+2}}{(W_1\cdot W_2)^{2s+2}},\notag\\
    &PT:(-1)^{2s}(c_e-ic_o)\frac{i^{-2s+2}}{(Z_1\cdot Z_2)^{-2s+2}}\to (c_e+ic_o)\frac{i^{-2s+2}}{(W_1\cdot W_2)^{-2s+2}}.
\end{align}
\subsubsection{Two point super-correlators}
Consider the two point integer spin super-correlator in the $(++)$ helicity. Its expression is,
\begin{align}
    (c_e+ic_o)\frac{i^{2s+2}}{(Z_1\cdot Z_2-\psi_1\psi_2)^{2s+2}}=(c_e+ic_o)\bigg(\frac{i^{2s+2}}{(Z_1\cdot Z_2)^{2s+2}}-i\psi_1\psi_2(2s+2)\frac{i^{2s+3}}{(Z_1\cdot Z_2)^{2(s+\frac{1}{2})+2}}\bigg).
\end{align}
The $(--)$ helicity counterpart is given by,
\begin{align}
    (c_e-ic_o)(-1)^{2s}\frac{i^{2s+2}}{(W_1\cdot W_2+\Bar{\psi}_1\Bar{\psi}_2)^{2s+2}}=(c_e-ic_o)\bigg(\frac{(-1)^{2s}i^{2s+2}}{(W_1\cdot W_2)^{2s+2}}-i\Bar{\psi}_1\Bar{\psi}_2(2s+2)\frac{(-1)^{2(s+\frac{1}{2})}i^{2s+3}}{(W_1\cdot W_2)^{2(s+\frac{1}{2})+2}}\bigg).
\end{align}
By comparing the component correlators using the super field expansion, we see that they are indeed $CPT$ conjugate using \eqref{PTSH}. Now, if we demand the CPT invariance for the whole super-correlator, we find that the CPT conjugate of $\psi$ must be $i\Bar{\psi}$ as in \eqref{PTGrassmann}. A similar analysis can be carried out for three point functions but we do not present the explicit calculations here.

\section{From supertwistor to spinor-helicity}\label{app:stsh}
We enumerate the results of $\mathcal{N}=1$ case in spinor-helicity variables in this appendix. They are obtained either by solving the supersymmetric Ward identity \eqref{QSH} or by performing an inverse half-Fourier transform on supertwistor correlators. We choose to work in the $\xi,\Bar{\xi}$ rather than the $\psi,\Bar{\psi}$ variables to allow easy comparison with the earlier work \cite{Jain:2023idr}.

\subsection*{Two point correlators}
The two point function in plus helicity is as follows,
\begin{align}
\langle \mathbf{\hat{J}}_{s}^{+}\mathbf{\hat{J}}_{s}^{+}\rangle=-(c_e+ic_o)\frac{\langle\bar{1}\,\bar{2}\rangle^{2s+1}}{16 p_1^2}\bigg(\xi_{+1}\xi_{+2}-\frac{4\langle 1 2\rangle}{p_1}\bigg),
\end{align}
whereas its conjugate minus helicity is given by,
\begin{align}
\langle \mathbf{\hat{J}}_{s}^{-}\mathbf{\hat{J}}_{s}^{-}\rangle=(c_e-ic_o)\frac{\langle 1\,2\rangle^{2s+1}}{16 p_1^2}\bigg(\bar{\xi}_{+1}\bar{\xi}_{+2}+\frac{4\langle \bar{1} \bar{2}\rangle}{p_1}\bigg).
\end{align}

\subsection*{Three point correlators}
The eight helicity configurations are given by,
\begin{align}\label{newSSHvarresultsNeq1}
&\langle \mathbf{\hat{J}}_{s_1}^{+}\mathbf{\hat{J}}_{s_2}^{+}\mathbf{\hat{J}}_{s_3}^{+}\rangle=\langle J_{s_1}^{+}J_{s_2}^{+}J_{s_3}^{+}\rangle\bigg(1-\frac{1}{8E}\big(\langle \bar{1}\bar{2}\rangle\xi_{+1}\xi_{+2}+\langle \bar{2}\bar{3}\rangle \xi_{+2}\xi_{+3}+\langle\bar{3}\bar{1}\rangle\xi_{+3}\xi_{+1}\big)\bigg),\notag\\&\langle \mathbf{\hat{J}}_{s_1}^{-}\mathbf{\hat{J}}_{s_2}^{-}\mathbf{\hat{J}}_{s_3}^{-}\rangle=\langle J_{s_1}^{-}J_{s_2}^{-}J_{s_3}^{-}\rangle\bigg(1+\frac{1}{8E}\big(\langle 12\rangle\bar{\xi}_{+1}\bar{\xi}_{+2}+\langle 23\rangle \bar{\xi}_{+2}\bar{\xi}_{+3}+\langle 31\rangle\bar{\xi}_{+3}\bar{\xi}_{+1}\big)\bigg),\notag\\&\langle \mathbf{\hat{J}}_{s_1}^{+}\mathbf{\hat{J}}_{s_2}^{+}\mathbf{\hat{J}}_{s_3}^{-}\rangle=\langle J_{s_1}^{+}J_{s_2}^{+}J_{s_3}^{-}\rangle\bigg(1-\frac{1}{8(E-2p_3)}\big(\langle \bar{1}\bar{2}\rangle\xi_{+1}\xi_{+2}-\langle \bar{2}3\rangle\xi_{+2}\bar{\xi}_{+3}+\langle 3\bar{1}\rangle\bar{\xi}_{+3}\xi_{+1}\big)\bigg)\notag\\&
\langle \mathbf{\hat{J}}_{s_1}^{-}\mathbf{\hat{J}}_{s_2}^{-}\mathbf{\hat{J}}_{s_3}^{+}\rangle=\langle J_{s_1}^{-}J_{s_2}^{-}J_{s_3}^{+}\rangle\bigg(1+\frac{1}{8(E-2p_3)}\big(\langle 12\rangle\bar{\xi}_{+1}\bar{\xi}_{+2}+\langle 2\bar{3}\rangle\bar{\xi}_{+2}\xi_{+3}+\langle \bar{3}1\rangle\xi_{+3}\bar{\xi}_{+1}\big)\bigg),\notag\\
&\langle \mathbf{\hat{J}}_{s_1}^{+}\mathbf{\hat{J}}_{s_2}^{-}\mathbf{\hat{J}}_{s_3}^{+}\rangle=\langle J_{s_1}^{+}J_{s_2}^{-}J_{s_3}^{+}\rangle\bigg(1-\frac{1}{8(E-2p_2)}\big(\langle \bar{1}2\rangle\xi_{+1}\bar{\xi}_{+2}+\langle 2\bar{3}\rangle\bar{\xi}_{+2}\xi_{+3}+\langle \bar{3}\bar{1}\rangle\xi_{+1}\xi_{+3}\big)\bigg),\notag\\&
\langle \mathbf{\hat{J}}_{s_1}^{-}\mathbf{\hat{J}}_{s_2}^{+}\mathbf{\hat{J}}_{s_3}^{-}\rangle=\langle J_{s_1}^{-}J_{s_2}^{+}J_{s_3}^{-}\rangle\bigg(1+\frac{1}{8(E-2p_2)}\big(\langle 1\bar{2}\rangle\bar{\xi}_{+1}\xi_{+2}+\langle \bar{2}3\rangle\xi_{+2}\bar{\xi}_{+3}+\langle 31\rangle\bar{\xi}_{+1}\bar{\xi}_{+3}\big)\bigg),\notag
\end{align}
\begin{align}
&\langle \mathbf{\hat{J}}_{s_1}^{-}\mathbf{\hat{J}}_{s_2}^{+}\mathbf{\hat{J}}_{s_3}^{+}\rangle=\langle J_{s_1}^{-}J_{s_2}^{+}J_{s_3}^{+}\rangle\bigg(1-\frac{1}{8(E-2p_1)}\big(\langle 1\bar{2}\rangle\bar{\xi}_{+1}\xi_{+2}+\langle \bar{2}\bar{3}\rangle\xi_{+2}\xi_{+3}+\langle\bar{3}1\rangle\xi_{+3}\bar{\xi}_{+1}\big)\bigg),\notag\\&
\langle \mathbf{\hat{J}}_{s_1}^{+}\mathbf{\hat{J}}_{s_2}^{-}\mathbf{\hat{J}}_{s_3}^{-}\rangle=\langle J_{s_1}^{+}J_{s_2}^{-}J_{s_3}^{-}\rangle\bigg(1+\frac{1}{8(E-2p_1)}\big(\langle\bar{1}2\rangle\xi_{+1}\bar{\xi}_{+2}+\langle 2 3\rangle\bar{\xi}_{+2}\bar{\xi}_{+3}+\langle 3\bar{1}\rangle\bar{\xi}_{+3}\xi_{+1}\big)\bigg).
\end{align}
The Euclidean space component correlators that multiply the Grassmann structures can be obtained via free theory computations via,
\begin{align}
\langle J_{s_1}^{h_1}J_{s_2}^{h_2}J_{s_3}^{h_3}\rangle=\langle J_{s_1}^{h_1}J_{s_2}^{h_2}J_{s_3}^{h_3}\rangle_{F}-\langle J_{s_1}^{h_1}J_{s_2}^{h_2}J_{s_3}^{h_3}\rangle_{B},
\end{align}
where the subscript $F$ ($B$) stands for the correlator in the free-fermionic (bosonic) theory. When the spin triangle inequality is satisfied\footnote{The spin triangle inequality is $s_i+s_j\ge s_k\forall i,j,k\in\{1,2,3\},i\ne j\ne k$.}, all the mixed helicity configurations vanish and the $(---)$ and $(+++)$ correlators take the following simple form,
 \begin{align}\label{homogeneousmmmppp}
\langle J_{s_1}^{-}J_{s_2}^{-}J_{s_3}^{-}\rangle&=(c_{even}+i c_{odd})\frac{\langle 1 2\rangle^{s_1+s_2-s_3}\langle 2 3\rangle^{s_2+s_3-s_1}\langle 3 1\rangle^{s_1+s_3-s_2}}{E^{s_1+s_2+s_3}}p_1^{s_1-1}p_2^{s_2-1}p_3^{s_3-1},\notag\\
\langle J_{s_1}^{+}J_{s_2}^{+}J_{s_3}^{+}\rangle&=(c_{even}-i c_{odd})\frac{\langle \bar{1} \bar{2}\rangle^{s_1+s_2-s_3}\langle \bar{2} \bar{3}\rangle^{s_2+s_3-s_1}\langle \bar{3} \bar{1}\rangle^{s_1+s_3-s_2}}{E^{s_1+s_2+s_3}}p_1^{s_1-1}p_2^{s_2-1}p_3^{s_3-1},
 \end{align}
 where $c_{even}$ and $c_{odd}$ are the OPE coefficients corresponding to the parity even homogeneous and parity odd structures respectively. The corresponding Wightman functions can be obtained by performing Wick rotation \cite{Baumann:2024ttn,Bala:2025gmz}.

\section{Tensor decomposition for extended supersymmetry}\label{app:tensor}
In this Appendix, we present some useful tensor structure decompositions of R-symmetry indices for component correlators in super-correlator that are helpful for extended supersymmetry.  Using supercurrent equation \eqref{supertwistorcurrentExt} for $\mathcal{N}= 2,3,4$ we get the component correlators with numerous decompositions.\\
For $\mathbf{\mathcal{N}=2}$, the following tensor structures are used,
\begin{align*}
\langle J_{s_1}^{A}J_{s_2}\rangle = 0,&\qquad
\langle J^{A}_{s_1}J^{B}_{s_2}\rangle = \delta^{AB}\langle J_{s_1}J_{s_2}\rangle,\\
\langle J^{A}_{s_1}J_{s_2}J_{s_3}\rangle= 0, \qquad
\langle J^{A}_{s_1}J^{B}_{s_2}J_{s_3}\rangle = (\delta^{AB}&+k \epsilon^{AB})\langle J_{s_1}J_{s_2}J_{s_3}\rangle,\qquad
\langle J^{A}_{s_1}J^{B}_{s_2}J^{C}_{s_3}\rangle =0.
\end{align*}
For $\mathbf{\mathcal{N}=3}$, the following tensor structures are used,
\begin{align*}
\langle J_{s_1}^{A}J_{s_2}\rangle &= 0,\qquad
\langle J^{A}_{s_1}J^{B}_{s_2}\rangle = \delta^{AB}\langle J_{s_1}J_{s_2}\rangle,\\
\langle J^{A}_{s_1}J_{s_2}J_{s_3}\rangle= 0,\qquad
\langle J^{A}_{s_1}J^{B}_{s_2}J_{s_3}\rangle &= \delta^{AB}\langle J_{s_1}J_{s_2}J_{s_3}\rangle,\qquad
\langle J^{A}_{s_1}J^{B}_{s_2}J^{C}_{s_3}\rangle = \epsilon^{ABC}\langle J_{s_1}J_{s_2}J_{s_3}\rangle.
\end{align*}
For $\mathbf{\mathcal{N}=4}$ the following tensor structures are used,
\begin{align*}
\langle J_{s_1}^{A}J_{s_2}\rangle = 0,\qquad
\langle J_{s_1}^{AB}J_{s_2}^{C}\rangle = 0&,\\
\langle J_{s_1}^{AB}J_{s_2}\rangle =0,\qquad
\langle J_{s_1}^{A}J_{s_2}^{B}\rangle =\delta^{AB}\langle J_{s_1}J_{s_2}\rangle,\qquad
\langle J_{s_1}^{AB}J_{s_2}^{CD}\rangle = &(\delta^{AC}\delta^{BD} - \delta^{AD}\delta^{BC} + k \epsilon^{ABCD})\langle J_{s_1}J_{s_2}\rangle.
\end{align*}
\begin{align*}
\langle J^{A}_{s_1}J_{s_2}J_{s_3}\rangle= 0, \qquad
\langle J^{A}_{s_1}J^{B}_{s_2}J_{s_3}\rangle= \delta^{AB}\langle J_{s_1}J_{s_2}J_{s_3}\rangle, \qquad
\langle J^{AB}_{s_1}J_{s_2}J_{s_3}\rangle= 0, \qquad \langle J^{A}_{s_1}J^{B}_{s_2}J^{C}_{s_3}\rangle= 0, \\
\langle J^{AB}_{s_1}J^{C}_{s_2}J_{s_3}\rangle= 0,\qquad
\langle J^{AB}_{s_1}J^{C}_{s_2}J^{D}_{s_3}\rangle= (\delta^{AC}\delta^{BD} - \delta^{AD}\delta^{BC})\langle J_{s_1}J_{s_2}J_{s_3}\rangle, \qquad \langle J^{AB}_{s_1}J^{CD}_{s_2}J^{E}_{s_3}\rangle&= 0, 
\end{align*}
\begin{align*}
\langle J^{AB}_{s_1}J^{CD}_{s_2}J_{s_3}\rangle&= (\delta^{AC}\delta^{BD} - \delta^{AD}\delta^{BC} + k \epsilon^{ABCD})\langle J_{s_1}J_{s_2}J_{s_3}\rangle,\\
\langle J^{AB}_{s_1}J^{CD}_{s_2}J^{EF}_{s_3}\rangle&=((\delta^{AC}\delta^{DE}\delta^{FB}-\delta^{AF}\delta^{DE}\delta^{BC}-\delta^{AD}\delta^{CE}\delta^{FB}+\delta^{AF}\delta^{EC}\delta^{BD}\notag\\
&-\delta^{AC}\delta^{DF}\delta^{EB}+\delta^{AE}\delta^{FD}\delta^{CB}+\delta^{AD}\delta^{CF}\delta^{EB}-\delta^{AE}\delta^{FC}\delta^{BD}) \notag\\
& + k (\delta^{AC}\epsilon^{BDEF}-\delta^{AD}\epsilon^{BCEF}+\delta^{AE}\epsilon^{BCDF}-\delta^{AF}\epsilon^{BCDE}-\delta^{BC}\epsilon^{ADEF}\notag\\
&+\delta^{BD}\epsilon^{ACEF}-\delta^{BE}\epsilon^{ACDF}+\delta^{BF}\epsilon^{ACDE}-\delta^{CE}\epsilon^{ABDF}+\delta^{CF}\epsilon^{ABDE}\notag\\
&+\delta^{DE}\epsilon^{ABCF}-\delta^{DF}\epsilon^{ABCE}))\langle J_{s_1}J_{s_2}J_{s_3}\rangle.
\end{align*}

\bibliographystyle{JHEP}
\bibliography{biblio}
\end{document}